\PassOptionsToPackage{noend}{algpseudocode}

\documentclass[sigconf]{acmart}

\acmPrice{15.00}
\copyrightyear{2023}
\acmYear{2023}
\setcopyright{rightsretained}
\acmConference[CHI '23]{Proceedings of the 2023 CHI Conference on Human Factors in Computing Systems}{April 23--28, 2023}{Hamburg, Germany}
\acmBooktitle{Proceedings of the 2023 CHI Conference on Human Factors in Computing Systems (CHI '23), April 23--28, 2023, Hamburg, Germany}\acmDOI{10.1145/3544548.3581477}
\acmISBN{978-1-4503-9421-5/23/04}

\usepackage[utf8]{inputenc}
\usepackage{enumitem}
\usepackage{subfig}
\usepackage{url}
\usepackage{multirow}
\usepackage{graphicx}

\usepackage{savesym}
\savesymbol{checkmark}
\DeclareFontFamily{U}{bbding}{}
\DeclareFontShape{U}{bbding}{m}{n}{<-> bbding10}{}
\newcommand{\leftpointright}{\lower2pt\hbox{\fontencoding{U}\fontfamily{bbding}\selectfont\symbol{'021}}}

\usepackage{caption}
\usepackage{listings}
\usepackage{xcolor}
\colorlet{punct}{red!60!black}
\definecolor{background}{HTML}{EEEEEE}
\definecolor{delim}{RGB}{20,105,176}
\colorlet{numb}{magenta!60!black}

\lstdefinelanguage{json}{
    basicstyle=\scriptsize\ttfamily,
    numbers=left,
    numberstyle=\scriptsize,
    stepnumber=1,
    numbersep=8pt,
    showstringspaces=false,
    breaklines=true,
    frame=lines,
    backgroundcolor=\color{background},
    literate=
     *{0}{{{\color{numb}0}}}{1}
      {1}{{{\color{numb}1}}}{1}
      {2}{{{\color{numb}2}}}{1}
      {3}{{{\color{numb}3}}}{1}
      {4}{{{\color{numb}4}}}{1}
      {5}{{{\color{numb}5}}}{1}
      {6}{{{\color{numb}6}}}{1}
      {7}{{{\color{numb}7}}}{1}
      {8}{{{\color{numb}8}}}{1}
      {9}{{{\color{numb}9}}}{1}
      {:}{{{\color{punct}{:}}}}{1}
      {,}{{{\color{punct}{,}}}}{1}
      {\{}{{{\color{delim}{\{}}}}{1}
      {\}}{{{\color{delim}{\}}}}}{1}
      {[}{{{\color{delim}{[}}}}{1}
      {]}{{{\color{delim}{]}}}}{1},
}

\usepackage{researchpack}

\newcommand{\calvar}[1]{\ensuremath{\mathcal{#1}}}

\newcommand{\calP}{\calvar{P}}

\newcommand{\calR}{\calvar{R}}

\newcommand{\vecvar}[1]{\ensuremath{\boldsymbol{#1}}}

\newcommand{\vp}{\vecvar{p}}

\algrenewcommand{\algorithmicrequire}{\textbf{Precondition:}}
\algrenewcommand{\algorithmicensure}{\textbf{Output:}}

\usepackage{acmart-taps}

\begin{document}

\title{The Elements of Visual Art Recommendation}
\subtitle{Learning Latent Semantic Representations of Paintings}

\author{Bereket A. Yilma}
\author{Luis A. Leiva}
\email{name.surname@uni.lu}
\affiliation{%
  \institution{University of Luxembourg}
  \country{Luxembourg}
}

\begin{abstract} 
Artwork recommendation is challenging because it requires understanding how users interact with highly subjective content, the complexity of the concepts embedded within the artwork, and the emotional and cognitive reflections they may trigger in users.
In this paper, we focus on efficiently capturing the elements (i.e., latent semantic relationships) of visual art for personalized recommendation.  We propose and study recommender systems based on textual and visual feature learning techniques, as well as their combinations. We then perform a small-scale and a large-scale user-centric evaluation of the quality of the recommendations.
Our results indicate that textual features compare favourably with visual ones, whereas a fusion of both captures the most suitable hidden semantic relationships for artwork recommendation. Ultimately, this paper contributes to our understanding of how to deliver content that suitably matches the user's interests and how they are perceived.
\end{abstract}

\begin{teaserfigure}
\includegraphics[width=\textwidth]{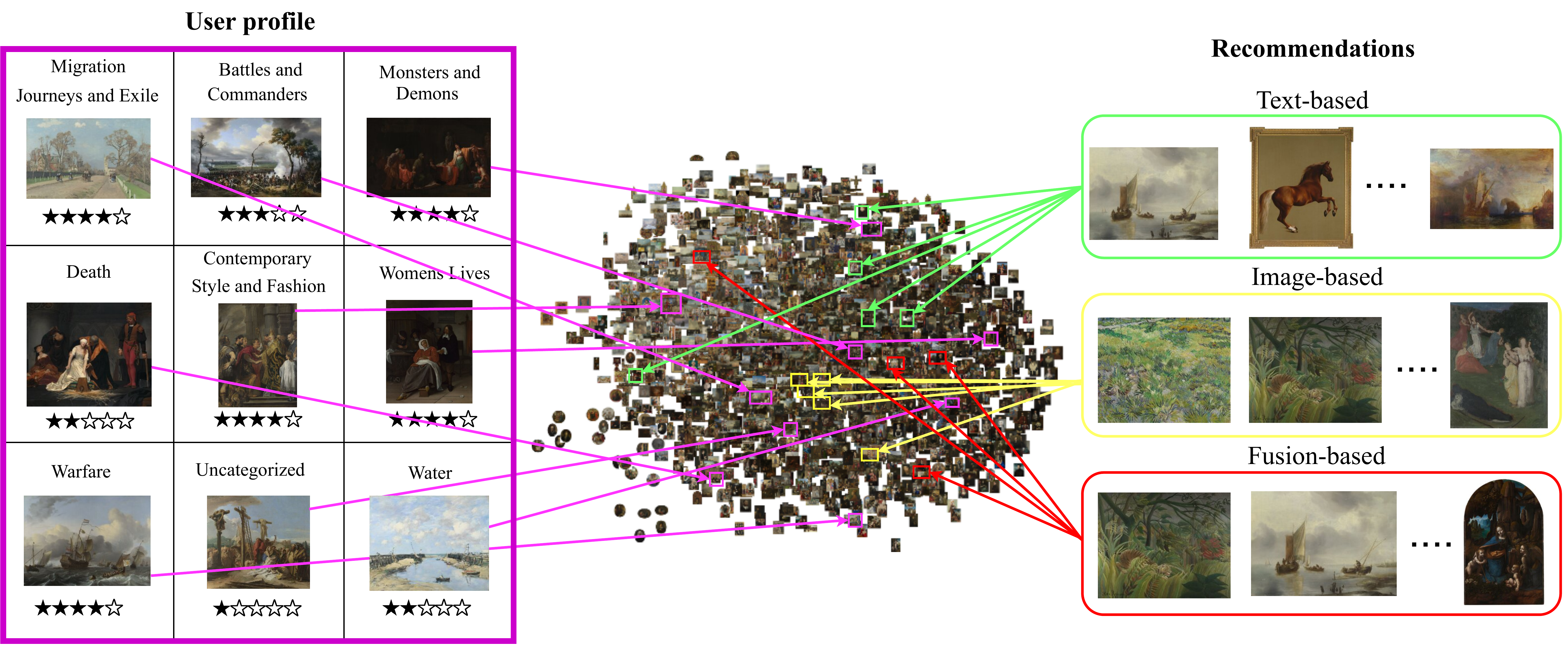}
\caption{Overview of personalized visual art recommendation.
    Based on a small set of elicited preferences (left),
    the user is shown different paintings (right) according to different recommender systems.}
\label{fig:LDA_plate}
\end{teaserfigure}



\maketitle

\section{Introduction}

In recent years, technology-mediated personalization and content recommendation has been areas of interest in Cultural Heritage environments  such as museums, art galleries, and exhibitions~\cite{naudet2015museum}. Although, in many cases the primary motivation for designing personalized services and recommender systems (RecSys) remains tightly linked to extrinsic motivation goals,
such as maximizing revenue, increasing user engagement, and optimizing advertisement delivery.
This approach to personalization may potentially overlook the very purpose of the cultural institutions
as well as the users' quality of experience~\cite{tsiropoulou2017quality},
who typically do it for their own pleasure, i.e., intrinsic motivation goals.
Thus, to enhance the perceived utility of RecSys,
it is of paramount importance to emphasize visitors’ quality of experience.
In this context, Visual Art (VA) recommendation is among the areas that has recently gained momentum~\citep{10.1145/3450613.3456847}.
Nevertheless, contrary to other application areas of RecSys where  personalised content is delivered to users
such as movies, music, news, etc., the domain of VA recommendation has not yet been sufficiently explored.

In the VA domain, paintings are important items that bring together complex elements
such as drawings, gestures, narration, composition, or abstraction~\cite{mayer1991artist}.
The task of personalized VA recommendation essentially entails suggesting paintings
that are similar to what a user has already seen or previously expressed interest.
The subjective nature of user's taste and the unique nature of their preferences,
which are long-standing challenges in content personalization,
are also salient issues in VA RecSys.
Especially since paintings carry deeper semantics than their traditional metadata,
i.e., categorizations based on their time period, technique, material, color, size, etc.
Furthermore, the kind of emotional and cognitive reflections paintings may trigger in users are also diverse, depending on their background, knowledge, and several other environmental factors~\cite{naudet2015museum}. Hence, to enhance personalized VA recommendations, efficiently capturing latent semantic relationships of paintings is vital and yet remains an open research challenge.

Most VA RecSys usually infer similarities and relationships among paintings from high-level features
derived from the above-mentioned traditional metadata such as artist names, styles, materials, and so on.
However, these features may not be expressive enough to capture abstract concepts
that are hidden in paintings and that could better adapt the recommendations to the subjective taste of the users.
For this, a high-quality representation of the data is crucial~\cite{bengio2013representation}.
Unfortunately, research on machine-generated data representation techniques for VA RecSys has been often overlooked,
as prominent works have largely relied on manually curated metadata~\cite{lykourentzou2013improving}.

\enlargethispage{12pt}
Recent work has started to pay more attention to machine-generated data representations to drive better VA recommendations.
He et al.~\cite{he2016vista} were among the first ones to use latent visual features
extracted using Deep Neural Networks (DNN) and also use pre-trained DNN models for VA recommendation.
A study reported by Messina et al.~\cite{messina2019content} showed that DNN-based visual features perform better
than leveraging textual metadata for VA recommendations.
However, they were focused on the artwork market,
which is driven by transaction data rather than enhancing the users' quality of experience.
Therefore, it is unclear if their findings would transfer to a more \textbf{user-centric} setting,
 which essentially entails investigating the actual relevance of recommendations to users
in terms of accuracy, novelty, diversity and serendipity.
Furthermore, they did not explore the combination of visual and textual features.

Alternatively, Yilma et al.~\cite{10.1145/3450613.3456847} proposed an approach to learn latent visual and textual features from paintings.
Their study indicated that recommendations derived from textual features compare favorably with visual ones.
Nonetheless, they also did not test hybrid approaches,
therefore it remains unclear which data representation technique (text, image, or a combination of both) is more efficient
to best capture the \emph{elements} (i.e., latent abstract concepts) embedded within visual arts for recommendation tasks. A recent work by Liu et al.~\cite{liu2020dynamic} have shown the benefit of jointly exploiting textual and visual features for recommendations. However, this has not been tested in the domain of VA RecSys.
To this end, we set out to explore techniques to learn latent semantic representation of paintings for personalized VA RecSys,
including the combination of each individual technique.

\enlargethispage{10pt}

Overall, previous works showed that visual features tend to perform better than textual metadata~\cite{messina2017exploring, 10.1145/3450613.3456847}
and hence they argued for not considering text-based information in VA RecSys.
In addition, it has not been explored yet whether hybrid approaches may yield better performance on VA recommendation tasks.
Therefore, we formulate the following research hypotheses:
\begin{description}
  \item[H1:] Visual features result in higher-quality recommendations than textual features.
  \item[H2:] Fusion of visual and textual features result in higher-quality recommendations than either could individually.
\end{description}

The first hypothesis is aimed at re-assessing our current understanding of the state of the art in VA RecSys research, whereas the second one, to the best of our knowledge, has never been assessed before in the domain of VA RecSys.

In this paper, we propose three different latent feature learning techniques leveraging both textual descriptions and images of paintings.
To learn latent features from textual descriptions, we adopt Latent Dirichlet Allocation (LDA)~\cite{blei2003latent} and Bidirectional Encoder Representations from Transformers (BERT)~\cite{devlin2018bert},
whereas for visual feature learning we use the popular Residual Neural Network (ResNet)~\cite{he2016deep}.
We also adopt a late fusion strategy proposed by Cormack et al.~\cite{10.1145/1571941.1572114}
which allows to combine different ranking techniques for information retrieval.
We then conduct a small-scale and a large-scale study based on a user-centric evaluation framework~\cite{pu2011user}. Specifically, we evaluated how accurate, diverse, novel, and serendipitous were the generated recommendations for the users
and derive valuable guidelines from our findings.
In sum, this paper makes the following contributions:
\begin{itemize}
    \item We develop and study five VA RecSys engines: LDA, BERT, ResNet, and their combinations.
    \item We conduct a small-scale ($N=11$) and a large-scale study ($N=100$)
        to assess VA RecSys performance from a user-centric perspective.
    \item We contextualize our findings and provide guidance about how to design next-generation VA RecSys.
\end{itemize}

\section{Related work}
\label{sec:related-work}

RecSys are becoming more and more prevalent in Cultural Heritage environments such as museums and art galleries~\cite{kontiza2018museum}.
The huge potential and benefit of personalized recommendations, in particular in the field of visual arts,
has been discussed by Esman~\cite{esman2012world}.
In the following we review previous work on VA recommendation and feature learning approaches.

\subsection{Recommending paintings}

According to Falk et al.~\cite{falk2016identity} the main motivation of museum visitors
is to have fun, experience art, learn new things, feel inspired, and interact with others.
When using digital museum guides, visitors' expectations are not only to be exposed to artwork that matches their interest
but also learn more and have access to more information~\cite{helal2013lessons}.

Research studies such as the CHIP project~\cite{aroyo2007personalized},
which implemented a RecSys for Rijksmuseum,\footnote{https://www.rijksmuseum.nl/en}
demonstrated the potential of personalization in such environments.
Hence, over the years, different kinds of RecSys have been exploited to provide personalized experiences to museum visitors.
For example, Aroyo et al.~\cite{hage2010finding} proposed a semantically-driven RecSys
and semi-automatic generation of personalized museum visits guided by visitor models.
Deladiennee et al.~\cite{8022674} introduced a graph-based semantic RecSys
that relies on an ontological formalisation of knowledge about manipulated entities.
Similarly, Kuflik et al.~\cite{kuflik2014graph} highlighted the benefits of graph-based recommendations.
This work was based on the premise that parts of the underlying data in a museum context
can be represented naturally by a graph that consists of typed entities and relations.
On the contrary, Frost et al.~\cite{frost2019art} introduced an anti-recommendation approach called \textit{``Art I don’t like''}
which exposes users to a variety of content and suggests artworks that are dissimilar to the ones the users selected,
aiming to maximize serendipity and exploration.
This method provides content that is aesthetically related in terms of low-level features,
but challenges the implied conceptual frameworks, which are driven by the preferences elicited by the users.
The very notion of this work was inspired by the work of Pariser~\cite{pariser2011filter}
which states that removing access to opposing viewpoints can lead to \textit{filter bubbles} in personalization.
Pariser's idea describes a type of ``intellectual isolation'' issue that occurs as a result of personalization algorithms.
These algorithms typically offer information to users that match previously viewed content and content viewed by similar users.
Hence, users have little exposure to contradicting viewpoints and become unknowingly trapped in a digital bubble.
This is a long-standing issue in RecSys and the community has explored different approaches to mitigate it,
e.g. improving transparency by giving the user control over the settings of the personalization algorithms~\cite{cosley2003seeing, cacheda2011comparison} and making recommendations understandable to users~\cite{forbes2012visualizing}.
However, there are several aspects that remain challenging in VA RecSys.
Primarily, because paintings are both high-dimensional and semantically complex,
we need a computationally efficient way of modelling both their content and their context. This essentially calls for efficient data representation techniques that are capable of capturing the complex semantics embedded in paintings.
Secondly, it also demands a more accurate representation of user profiles such as modelling temporal and social dynamics in terms of users' tendency to interact with content more or less consistently,
as well as their preferences towards individual artists, styles, colors, etc.
However, these are rarely available or not directly accessible in practice, making the so called cold-start problem\footnote{When the system has no information about the users, it cannot provide personalised recommendations.} a prevalent issue in VA RecSys.

\subsection{Learning painting features}

He et al.~\cite{he2016vista} proposed a visually, socially, and temporally-aware model for artistic recommendation.
This was among the first works that utilized the power of DNNs to exploit latent representations for VA recommendation.
Their work primarily builds upon two methods, factorized personalized Markov chains (FPMC)~\cite{rendle2010factorizing}
and visual Bayesian personalized ranking (VBPR)~\cite{he2016vbpr}.
On the one hand, FPMC was adopted to capture the fact that users tend to browse art with consistent latent attributes
during the course of a browsing session, as FPMC models the notion of smoothness between subsequent interactions using a Markov chain.
On the other hand, VBPR models the visual appearance of the items being considered.
By combining the two models, He et al. tried to  capture individual users' preferences towards particular VA styles,
as well as the tendency of users to interact with items that are `visually consistent' during a browsing session.
They also proposed several extensions of these models to handle longer memory than simply previous actions.
Unfortunately, their method is only applicable under the collaborative filtering scenario,
for example matching products to users based on past purchases.
However, collaborative filtering suffers from the above-mentioned cold-start problem.
In addition, they did not investigate explicit visual features nor textual metadata.

Subsequently, Messina et al.~\cite{messina2017exploring, messina2019content, messina2020curatornet} explored content-based artwork recommendation using images, keywords, and transaction data from the UGallery online artwork store.\footnote{\url{https://www.ugallery.com/}}
Their work suggested that automatically computed visual features perform better than manually-engineered visual features extracted from images (i.e, texture, sharpness, brightness, etc.). Their work also indicated that a hybrid approach combining visual features and textual keyword attributes such as artist, title, style, etc., yields a further performance improvement. However, their hybrid approach was based on computing a score as a convex linear combination of the scores of individual methods (visual similarity and keyword similarity). Particularly, they did not explore feature learning approaches
such as topic modeling techniques we study in this paper, which are more scalable and generalizable.
Furthermore, their work was focused on predicting future purchases of artwork rather than enhancing personal experiences.

Recent works by Yilma et al.~\cite{10.1145/3450613.3456847, yilma2020personalised} proposed a VA recommendation approach
that leveraged topic modeling techniques from textual descriptions of paintings
and performed a comparative study against visual features automatically extracted using DNNs.
Their study demonstrated the potential of learning features from text-based data,
especially when it comes to explaining the recommendations to the user.
However, they never looked at the combination of text-based and image-based RecSys engines.

In sum, a number of VA Recsys strategies have been proposed over the years,
but given that (i)~user preferences are highly subjective and (ii)~visual artwork is particularly complex to grasp,
VA recommendation remains a rather challenging task.
Thus, research effort in uncovering latent semantics of visual art is still considered a worthwhile endeavour,
especially with regards to evaluating the quality of the recommendations from a user-centric perspective.
To the best of our knowledge, this paper is the first to systematically shed light in this regard.

\section{Background: Learning latent representations of paintings}

\begin{figure*}[!h]
\centering
\includegraphics[width=\textwidth]{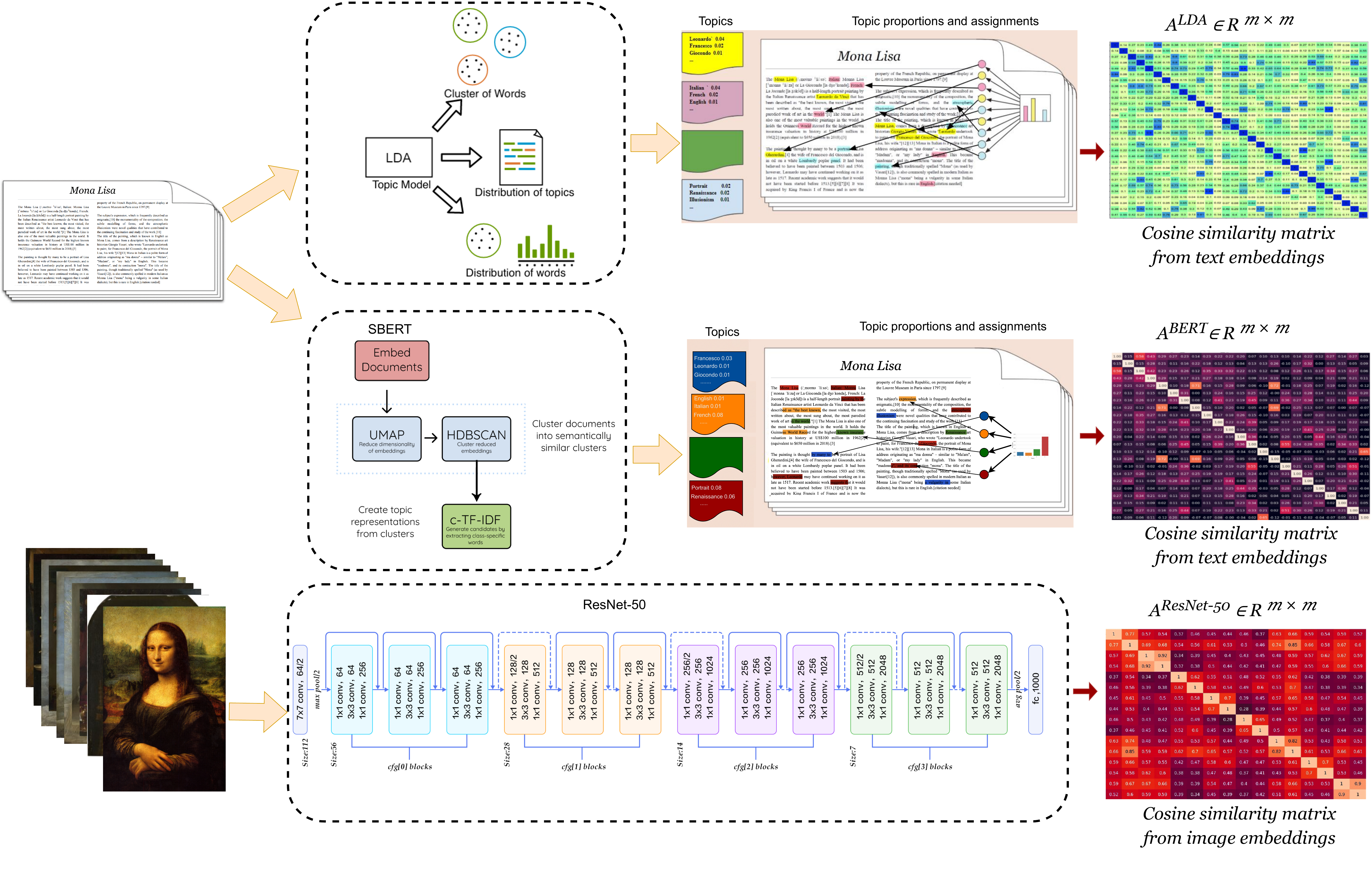}
\caption{\textbf{The elements of VA recommendation}: Overview of our approaches to learn latent semantic representations of paintings.}
\label{fig:feature_learning}
\end{figure*}

Data representation techniques play a great role in VA RecSys,
as they can entangle and reveal interesting factors embedded within the artwork data,
thereby eventually influencing the quality of the recommendations~\cite{roy2022systematic}.
Specifically, the complexity of the concepts embodied within paintings
makes the task of capturing semantics by machines far from trivial.
To this end, we set out to study  different representation techniques
that can efficiently learn the elements (i.e., latent semantic relationships of paintings) of VA RecSys.
\autoref{fig:feature_learning} summarizes the three painting representation learning approaches we propose and study in this paper.

\subsection{Feature learning from Text-based representations of Paintings}

In Natural Language Processing and Information Retrieval,
vector space models have been used to represent documents efficiently~\cite{bengio2013representation}.
However, this kind of representations has a limited ability to capture inter/intra-document relationships.
It has been shown that, as data dimensionality increases,
the distance to the nearest data point approaches the distance to the furthest data point~\cite{Aggarwal01}.
Consequently, in high dimensional spaces the notion of spatial locality becomes ill-defined~\cite{Beyer99}.
Hence, researchers have been proposing more advanced techniques aiming to tackle the curse of dimensionality reduction
and to better capture hidden semantic structures in document modeling.
Among these efforts, Latent Dirichlet Allocation (LDA),
an unsupervised generative probabilistic model proposed by Blei et al.~\cite{blei2003latent},
has demonstrated superiority over several other models.

LDA has been applied in several text-based RecSys tasks
such as scientific paper recommendation~\cite{amami2016lda},
personalized hashtag recommendation~\cite{zhao2016personalized},
and online course recommendation~\cite{apaza2014online}, among others.
On the other hand, a more recent work by Devlin et al.~\cite{devlin2018bert}
developed Bidirectional Encoder Representations from Transformers (BERT)
and set a new state-of the-art performance on sentence-pair related tasks
like semantic textual similarity and question answering.
However, BERT entails an important computational overhead due to the many possible combinations for prediction.
For example, to find the most similar pairs in a collection of 10,000 sentences,
BERT requires about 50 million inference computations.
Sentence-BERT (SBERT)~\cite{DBLP:journals/corr/abs-1908-10084},
a modification of the pre-trained BERT model,
managed to reduce the computation time from 65 hours to 5 seconds in a single V-100 GPU.
It uses siamese and triplet network structures to derive semantically meaningful sentence embeddings
that can be compared using the tried-and-true cosine similarity measure.
In the past few years, BERT models have proven to be powerful document embedding techniques
and have been used for extracting latent semantic structures (i.e., topics)
underlying a large set of documents~\cite{grootendorst2022bertopic}.
As a result, BERT has gained tremendous popularity in the design of RecSys
that exploit textual data~\cite{hassan2019bert, juarto2021neural, lavi2021consultantbert}.

We adopt LDA and BERT (actually SBERT) to learn painting feature representations from their associated textual metadata,
where each painting is represented by a document containing detailed annotations such as title, format, or a curated description; see \autoref{fig:ng_sample} for an example. In essence, a painting can be described as a mixture of several concepts such as religion, nudity, portrait, etc.  Thus, each document is a distribution of topics and each topic is a distribution of words.
Prominent words in each latent topic explain the nature of the topic   and prominent latent topics related to each document explain the nature of the document (i.e. paintings).   For example, let us assume that latent topics are ``religion'', ``still life'', and ``landscape''.  A painting may have the following distribution over the topics: 70\% ``religion'', 10\% ``still life'', and 20\% ``landscape''.  Moreover, each topic has a distribution over the words in the vocabulary.  For the ``religion'' topic, the probability of the word ``Saint'' would be much higher than in the ``landscape'' topic.  Hence, the employed LDA and BERT representation techniques will find high-dimensional vector representations  that capture the topic proportions for each painting in such a way that semantically similar paintings are closer to each other in the feature representation space.
In the following we briefly discuss each text-based feature learning approaches in more detail.

\subsubsection{Latent Dirichlet allocation (LDA)}

LDA is an unsupervised learning algorithm that attempts to describe a set of observations as a mixture of distinct categories.
Each observation is a document, the features are the presence (or occurrence, or count) of words, and the categories are the topics.
The topics themselves are not specified up-front, only their number,
since they are learned as a probability distribution over the words that occur in each document. The procedure of building an LDA model for VA RecSys is described as follows.
We start by constructing a collection of documents containing textual information about each painting.
Then, a desired number of topics $k$  is chosen and a topic is attributed to each word $w$ in the collection of documents
where $\theta_{i} \sim Dir(\alpha)$; $\theta$ is the topic distribution for a document $d$ and $\alpha$ is the per-document topic distribution, with $i \in \{1, ..., k\}$ and $Dir(\alpha)$ is a Dirichlet distribution over the $k$ topics.
Subsequently, the learning is done by computing the conditional probabilities $P(t|d)$ (i.e., the likelihood of topic $t$ given document $d$) and $P(w|t)$ (i.e., likelihood of word $w$ given topic $t$). A detailed discussion on LDA topic modeling can be found in~\cite{blei2003latent} and~\cite{jelodar2019latent}.

Once the LDA model is trained over the entire text dataset, 
a matrix $\mathbf{A} \in \mathbb{R}^{m \times m}$ is produced where each entry $a(i,j)$
is the cosine similarity measure between document embeddings.
This similarity matrix therefore captures the latent topic distribution over all documents,
which is then leveraged to compute semantic similarities of paintings for VA RecSys tasks,
as explained in the next section.

\subsubsection{Bidirectional Encoder Representations from Transformers (BERT)}
The second approach we study to learn latent feature representations of paintings from their associated textual metadata
has been recently proposed by Grootendorst et al.~\cite{grootendorst2022bertopic} and is based on sentence Transformers.
Similar to the LDA approach,
we begin by constructing a collection of documents with textual metadata of paintings.
Then, feature learning is done in three steps.
First, each painting document is converted to an embedding representation
using the pre-trained SBERT large language model,\footnote{We used the ``all-MiniLM-L6-v2'' version, to optimize performance, but any other version can provide suitable painting embeddings.}
which maps sentences and paragraphs to a 384-dimensional dense vector space~\cite{DBLP:journals/corr/abs-1908-10084}.
Second, the dimensionality of the embeddings is reduced using the uniform manifold approximation and projection (UMAP) algorithm~\citep{mcinnes2018umap}.
This allows to learn a more efficient representation while at the same time preserving the global structure of the original embeddings.
Third, the reduced embeddings are semantically clustered together using HDBSCAN~\cite{campello2013density},
a soft-clustering algorithm that prevents unrelated documents to be assigned to any cluster.
Finally, latent topic representations are extracted from the clusters
using a custom class-based term frequency--inverse document frequency (c-TF-IDF) algorithm,
which produces importance scores for words within a topic cluster.
The main idea of c-TF-IDF is that extracting  the most important words per cluster yields descriptions of topics.
Hence, TF-IDF is adjusted and the inverse document frequency is replaced
by the inverse class frequency to measure how much information a term provides to a class.
Formally the c-TF-IDF of a word $w$ in class $C$ is given by:
\begin{equation}\label{eq:c-tf-idf}
  \text{c-TF-IDF}(w,C) = f_{w,C} \cdot \log(1 + \frac{N}{f_{w}})
\end{equation}
where $f_{w,C} = \frac{|w|}{\sum_{c \in C} |c|}$ is the frequency of word $w$ in class $C$,
$N$ is total number of words per class,
$f_{w}$ is the frequency of word $w$ across all classes,
and $| \cdot |$ denotes the number of items in a set.
Words with high c-TF-IDF scores are selected for each topic $t$,
thereby producing topic-word distributions for each cluster of documents $d$.

Once the BERT model is trained over the entire dataset,
a matrix $\mathbf{A} \in \mathbb{R}^{m \times m}$ is produced where each entry
is the cosine similarity measure between all document embeddings.
Again, this similarity matrix captures the latent topic distribution over all documents,
which is then leveraged to compute semantic similarities of paintings for VA RecSys tasks,
as explained in the next section.

\subsection{Feature learning from image-based representations of paintings}
\label{subsec:resnet}

Visual feature extraction is critical to have a discriminative representation of images~\cite{malik2022computer},
and it is widely used in several tasks such as object detection, classification, or segmentation~\cite{salau2019feature}.
Traditional approaches to feature extraction include Harris Corner Detection~\cite{chen2009comparison},
or the more advanced version Shi-Tomasi Corner Detector~\cite{bansal2021efficient}.
Other approaches have been proposed, such as SURF~\cite{mair2010adaptive} or BRIEF~\cite{calonder2010brief},
but they have been superseded by recent advances in Deep Learning, in particular in Convolutional Neural Networks (CNN).

Today, image feature extraction techniques are mostly based on pre-trained CNN architectures
such as AlexNet~\cite{krizhevsky2017imagenet}, GoogLeNet~\cite{szegedy2015going}, and VGG~\cite{simonyan2014very}.The winner of the 2015 ImageNet challenge, ResNet,
proposed by He et al.~\cite{he2016deep} introduced the use of residual layers to train very deep CNNs,
setting a world record of more than 100 layers. ResNet-50 is the 50-layer version of this architecture,
trained on more than a million images from the ImageNet database.\footnote{\url{http://www.image-net.org}}
Thus, it has learned rich feature representations for a wide range of images
and has shown superiority over other pre-trained models as a feature extractor~\cite{ikechukwu2021resnet,li2021facial,barata2018survey}.

We used the ResNet-50 model pre-trained on ImageNet to extract latent visual features (image embeddings) from paintings.
By passing each painting image through the network,
a convolutional feature map (i.e., a feature vector representation) is obtained.
Once we extract all image features from the entire dataset,
a matrix $\mathbf{A} \in \mathbb{R}^{m \times m}$ is produced where each entry
is the cosine similarity measure between all image embeddings.
This similarity matrix therefore captures the latent visual distribution over all images,
which is then leveraged to compute semantic similarities of paintings for VA RecSys tasks,
as explained in the next section.

\section{Method: Personalized recommendation of paintings}

We consider approaches that, together, can learn features from both textual and visual information from paintings.
We study two different techniques for learning text-based representations (LDA and BERT),
as there are no exhaustive prior works of VA RecSys leveraging textual data.
On the other hand, since visual features have been extensively explored in VA RecSys applications~\cite{messina2020curatornet,he2016vista, 10.1145/3450613.3456847, kelek2019painter}, we study ResNet-50 for learning image-based representations,
which it is considered the state of the art in prior work~\cite{10.1145/3450613.3456847, kelek2019painter}.
Let $P = \{p_1, p_2, \dots, p_m\}$ be a set of image paintings, %
$\calP = \{\vp_1, \vp_2, \dots, \vp_m\}$ be the associated embeddings of each painting
according to LDA, BERT, or ResNet,
and $P^u  = \{p^u_1, p^u_2, \dots, p^u_n\}$ be the set of paintings a user $u$ has rated,
where $P^u \subset P$ and $\omega^u=\{\omega^u_1, \omega^u_2, \dots, \omega^u_n\}$
are the normalized ratings that $u$ gave to a small set of paintings $P^u$.

Once the dataset embeddings (latent feature vectors) are learned using either model (LDA, BERT, or ResNet)
we compute the similarity matrix for all the paintings $\mathbf{A}$.
Next, the preferences of a user $u$ are modelled by a normalized vector
that transforms a simple 5-point scale rating into weights $\omega^{u}_i \in [0,1]$ for every painting $p^{u}_i$ the user has rated.
Then, the predicted score $S^u(p_i)$ the user would give to each painting in the collection $P$
is calculated based on the weighted average distance
between the rated paintings and all other paintings:
\begin{equation}\label{eq:user-score}
    S^u(p_i) =  \frac{1}{n} \sum^{n}_{j = 1} \omega^u_j \cdot \mathbf{A}_{ij}
\end{equation}
where $\mathbf{A}_{ij} = d(\vp_i, \vp_j)$ is the similarity between embeddings of paintings $p_i$ and $p_j$ in the computed similarity matrix. The summation in \autoref{eq:user-score} is taken over all user's rated paintings $n = |P^u|$.
Once the scoring procedure is complete, the paintings are sorted
and the $r$ most similar paintings constitute a ranked recommendation list.
In sum, the VA RecSys task consists of recommending the most similar paintings to a user
based on a small set of paintings rated before, i.e., the elicited preferences.

In this paper, we study five RecSys engines: three are based on LDA, BERT, and ResNet (non-fusion engines),
whereas the other two engines (fusion engines) are hybrid combinations (text+image) of the first three engines.
For the fusion engines, we adopted the ``reciprocal rank fusion'' strategy proposed by Cormack et al.~\cite{10.1145/1571941.1572114}
for combining rankings in information retrieval systems.
It is a late fusion technique that is easily composable and simple to use.
Late fusion (i.e. at post-hoc) is often preferred than early fusion (i.e. at the feature level)
because the models involved are independent from each other,
so each can use their own features, numbers of dimensions, etc.~\cite{Ramachandram17}.

Furthermore, with late fusion it is possible to precisely control the contribution of each model (e.g. 25\% text and 75\% image).
Although in our work both text and image features contribute equally (i.e., 50\% each) when it comes to producing the recommendations.
Our proposed VA RecSys engines are outlined in Algorithms 1 to 3, respectively.

\begin{algorithm*}[!ht]
\label{algo1}%
\caption{Dataset preprocessing.}
\begin{algorithmic}[1]
    \Procedure{Preprocess}{$\text{Model}, P$} \Comment{Model $\in \{\text{LDA, BERT, ResNet}\}$}
        \State $\calP \gets \Call{FeaturizePaintings}{\text{Model}, P}$ \Comment{Compute painting embeddings}
        \State $\mathbf{A} \gets \emptyset$ \Comment{Initialize $\mathbf{A} \in \mathbb{R}^{m \times m}$ and $m = |\calP|$}
        \For{$\vp_i$ and $\vp_j \in \calP$} \Comment{Iterate over collection}
            \State $\mathbf{A}_{ij} \gets$ \Call{CosineSimilarity}{$\vp_i, \vp_j$}
        \EndFor
        \State \Return $\mathbf{A}$
    \EndProcedure
\end{algorithmic}
\end{algorithm*}

\begin{algorithm*}[!ht]\setcounter{algorithm}{1}%
\label{algo2}%
\caption{
    Non-fusion VA RecSys.
}
\begin{algorithmic}[1]
    \Require Similarity matrix $\mathbf{A}$ of featurized paintings
    \Procedure{RecommendPaintings}{$\text{Model}, P^u, \omega^u, r$} \Comment{Model $\in \{\text{LDA, BERT, ResNet}\}$}
        \State $\calP^u \gets \Call{FeaturizePaintings}{\text{Model}, P^u}$ \Comment{Load computed embeddings, as $P^u \subset P$}
        \State $S^u \gets \emptyset$ \Comment{Initialize recommendations list}
        \For{$\vp_i \in \calP^u$ and $\vp_j \in \calP$} \Comment{Iterate over collection}
            \State $S^u(\vp_i) = \frac{1}{n} \sum^{n}_{j = 1} \omega^u_j \cdot \mathbf{A}_{ij}$ \Comment{Bias similarity matrix towards user's preferences}
        \EndFor
        \State \Call{Sort}{$S^u$} \Comment{Descending sort, as higher cosine distance means higher similarity}
        \State \Return \Call{Slice}{$S^u, r$} \Comment{Pick the first $r$ elements in the ranking}
    \EndProcedure
\end{algorithmic}
\end{algorithm*}

\begin{algorithm*}[!ht]\setcounter{algorithm}{2}
\label{algo3}%
\caption{Fusion-based VA RecSys.}
\begin{algorithmic}[1]
    \Procedure{FuseRecommendations}{$\text{Model}_1, \text{Model}_2, P^u, \omega^u, r$}
        \Comment{Models 1 \& 2 $\in \{\text{LDA, BERT, ResNet}\}$}
        \State $\calR_1 \gets$ \Call{RecommendPaintings}{$\text{Model}_1, P^u, \omega^u, r$} \Comment{Get recommended lists from each model}
        \State $\calR_2 \gets$ \Call{RecommendPaintings}{$\text{Model}_2, P^u, \omega^u, r$}
        \State $F(\vp \in \calR_1 \bigcup \calR_2) = \sum_{i \in \calR_1, j \in \calR_2} \frac{1}{n(i) n(j)}$ \Comment{Fuse lists: $n(\cdot)$ is the $n$th ranking position of each painting}
        \State \Call{Sort}{F} \Comment{Descending sort, as higher cosine distance means higher similarity}
        \State \Return \Call{Slice}{$F, r$} \Comment{Pick the first $r$ elements in the fused ranking}
    \EndProcedure
\end{algorithmic}
\end{algorithm*}

\section{Dataset}

\begin{figure*}[!ht]
\centering
\includegraphics[width= 0.9\textwidth]{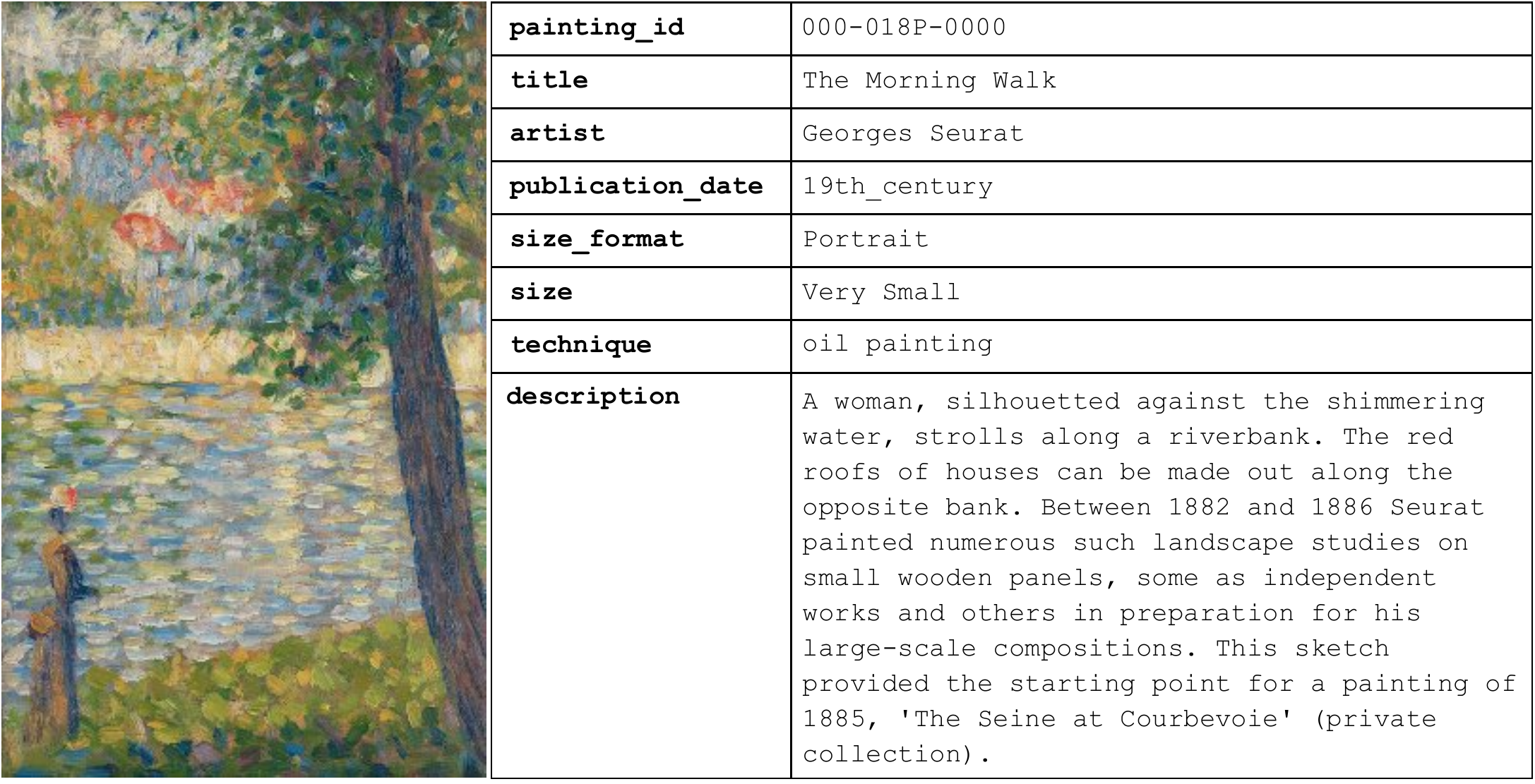}
\caption{Sample painting and associated metadata from the National Gallery dataset.}
\label{fig:ng_sample}
\end{figure*}

We used a dataset containing 2,368 paintings from The National Gallery, London.\footnote{https://www.nationalgallery.org.uk/}
This curated set of paintings belongs to the CrossCult Knowledge Base.\footnote{\url{https://www.crosscult.lu/}}
Each painting image is accompanied by a set of text-based metadata, which makes this dataset suitable for testing the proposed feature learning approaches. A sample data point is shown in \autoref{fig:ng_sample}.
For our text-based RecSys engines (LDA and BERT) we use all available painting attributes,
such as artist name, painting title, technique used, etc. as well as a description provided by museum curators.
These descriptions carry complementary information about the paintings
such as stories and narratives that can be exploited to better capture the painting semantics.
The image-based RecSys engine (ResNet) uses the convolutional feature maps automatically extracted from the painting images.

The dataset also provides curated stories that we study to sample initial user preferences in the profiling phase.
In the following subsections we present a detailed analysis of the dataset to better understand the behavior and implementation of our RecSys engines.

\subsection{Story groups}

\begin{figure*}[!h]
\centering
\includegraphics[width=\textwidth]{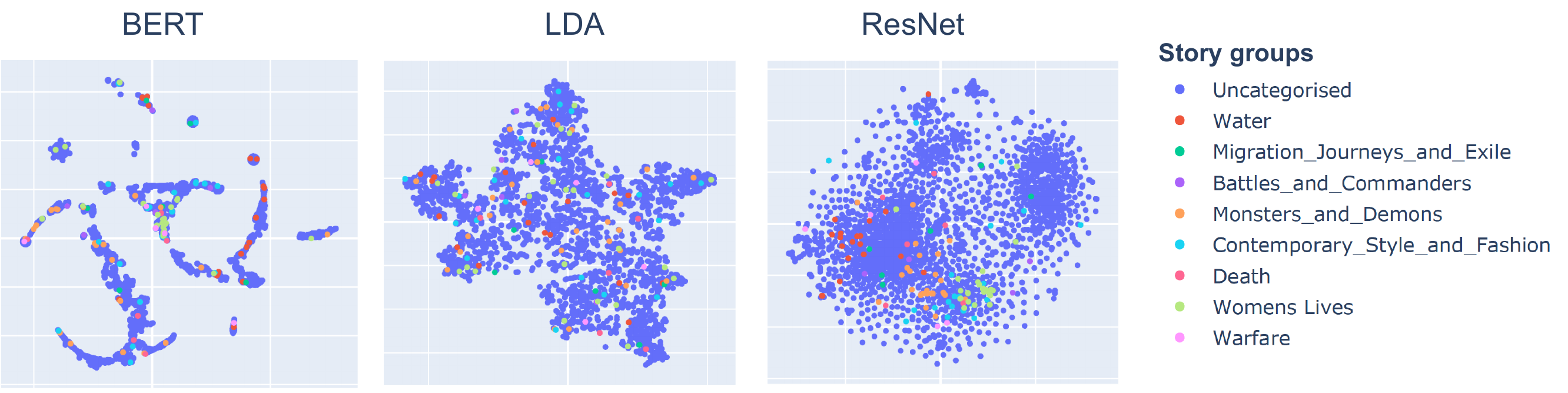}
\caption{Latent space projection (t-SNE) of the curated story groups.}
\label{fig:story-group}
\end{figure*}

The dataset provides 8 curated stories (categories) linked to a few of the paintings, namely:
\textit{`Women's lives', `Contemporary style and fashion', `Water, Monsters and Demons', `Migration: Journeys and exile', `Death', `Battles and Commanders', and `Warfare'}.
\autoref{fig:story-group} shows a 2D projection map of the story groups in the dataset
using the non-linear projection t-SNE algorithm~\cite{Maaten:2008:tSNE}.
We can see that the majority of the paintings belong to the `uncategorized' class.
These story groups are meant to provide context to a selected group of paintings,
according to the museum experts who created the dataset.
We can observe from the latent space projections that the story groups are scattered across the entire dataset,
suggesting that museum curators considered them to be representative examples of the collection.
The map projection also surfaces the complex latent semantic relationships among the paintings.

\enlargethispage{12pt}

\subsection{Preprocessing}

On the one hand, to learn textual features with LDA and BERT models,
the painting metadata were pre-processed:
text fields \hbox{concatenation}, removal of punctuation symbols and stop-words, lowercasing, and lemmatization.
On the other hand, to learn visual features with the ResNet model,\enlargethispage{12pt}\
we used the actual images of paintings\footnote{All paintings are available under a Creative Commons (CC) license.}
to extract the convolutional feature maps with the pre-trained ResNet-50 model discussed in \autoref{subsec:resnet}.

\subsection{Text source analysis}

In topic modeling, ``topic coherence" is a commonly used technique to evaluate topic models.
It is defined as the sum of pairwise similarity scores on the words $w_1, ..., w_n$
that describe each topic, usually the most frequent $n$ words according to $p(w|t)$~\cite{jelodar2019latent}:
\begin{equation}
 \textsc{TopicCoherence} = \sum ^{n}_{i < j} \textsc{CosineSimilarity}(w_i, w_j)
\end{equation}

\begin{figure}[!h]
\begin{centering}
\includegraphics[width= 0.4\textwidth]{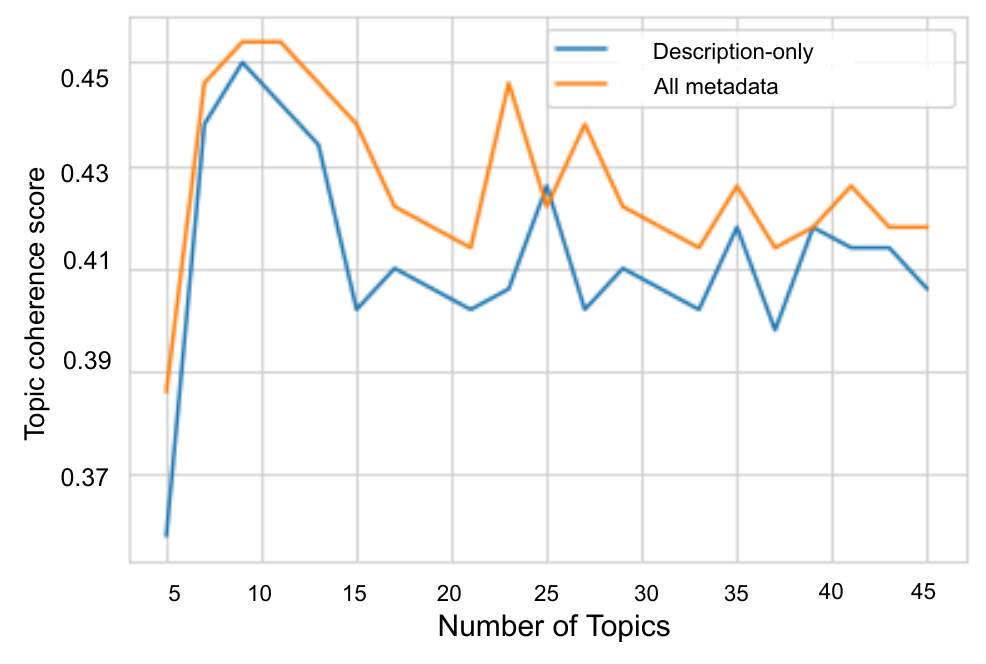}
\caption{Comparative topic coherence analysis of the two text sources.}
\label{fig:Top_COH}
\end{centering}
\end{figure}

Ideally, a good model should generate coherent topics;
i.e the higher the coherence score the better the model is~\cite{newman2010automatic}.
\autoref{fig:Top_COH} shows topic coherence in LDA as a function of the number of topics
when using two text sources:
`description-only' (using only the curated stories from the \texttt{description} metadata, see \autoref{fig:ng_sample})
and `all-metadata' (using all the available metadata shown in \autoref{fig:ng_sample}).

From the analysis presented in \autoref{fig:Top_COH} we can make two important observations.
First, a topic model using all the available metadata consistently gives better topic coherence scores,
therefore it is considered a better model.
Using only the curated stories is a close contender,
however we should note that it is very time consuming to produce these, as they require human expertise.
Second, the maximal topic coherence is obtained at 10 topics when using all the available metadata.
Having too many topics requires more resources as well as more computation time,
therefore it is important to find a reasonable balance.
On the other hand, for topic modelling using BERT we rely on HDSCAN to choose the optimal number of initial clusters
and then latent topics are determined based on c-TF-IDF over those clusters.

\begin{figure*}[!h]
\centering
\includegraphics[width= 0.7\textwidth]{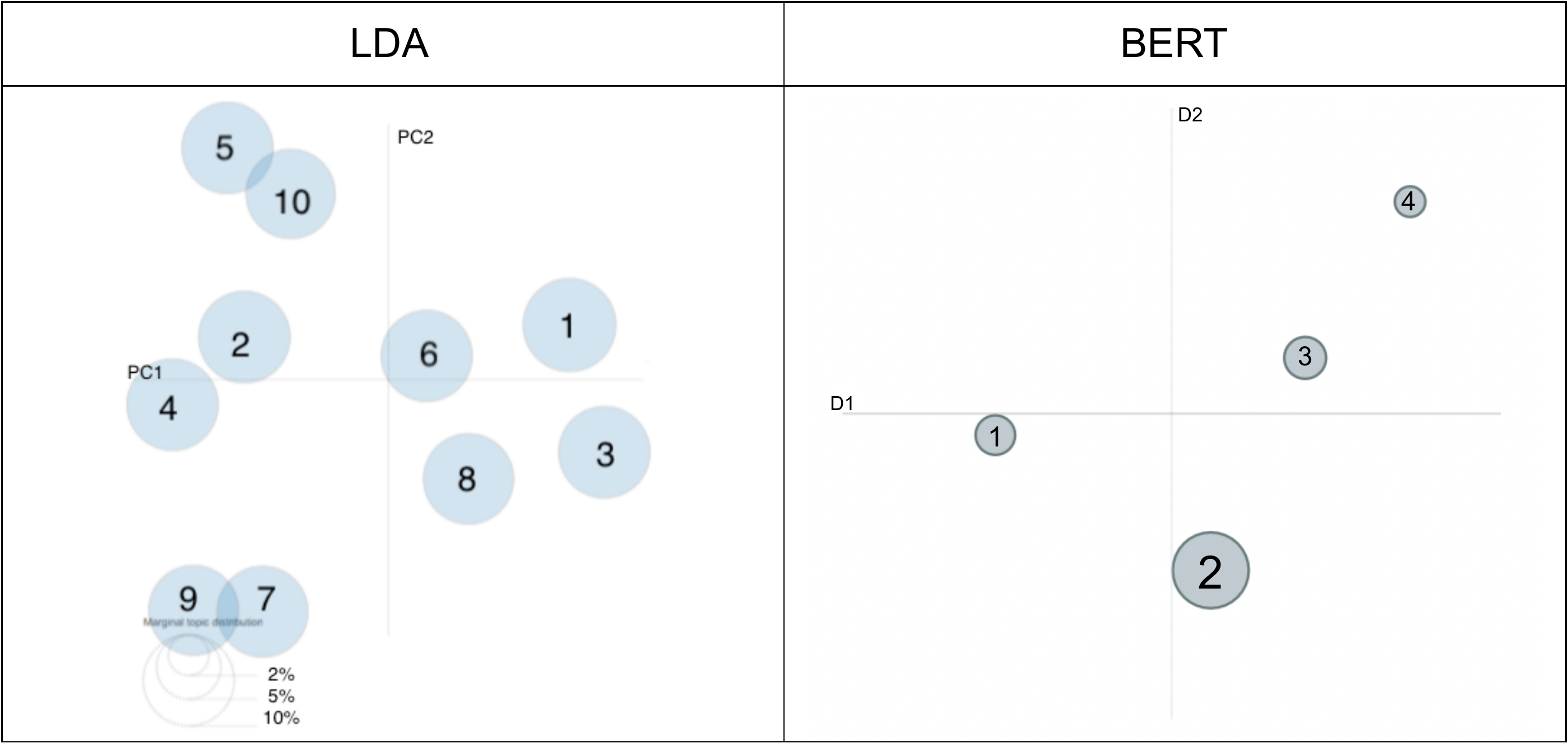}
\caption{ Inter-topic distance map of LDA and BERT in a projected 2-dimensional space.}
\label{fig:intertopic}
\end{figure*}

\autoref{fig:intertopic} shows the generated topics by LDA and BERT for our dataset.
The size of each circle represents the prevalence of a topic, i.e, the popularity of a topic among the paintings.
The distance between the circles represents the similarity between topics. The objective here is to have topics that are overlapping as little as possible.
For LDA, the 10 automatically identified topics are evenly popular
while being sufficiently distinct from each other with some overlaps between topic 9 \& 7 and 5 \& 10.
For BERT, the four automatically identified topics are significantly distinct from each other
while they are also laid out in descending order or popularity across the dataset.
This indicates that after clustering, similar topics are merged together
to reduce the total number of topics, thereby creating more cohesive topic models.

Hence, the number of topics in our implementation were set to 10 and 4 for LDA and BERT respectively.

\section{Evaluation}

The main goal of our evaluation was to understand the user's perception towards the quality of our studied RecSys engines,
and ultimately to assess which feature learning approach best captures the semantic relatedness of paintings.
We conducted two user studies, to be described later,
that were approved by the Ethics Review Panel of the University of Luxembourg with ID 22-031.

\subsection{Apparatus}

\begin{figure*}[!ht]
    \centering

    \def\h{6cm}
    \includegraphics[height=\h]{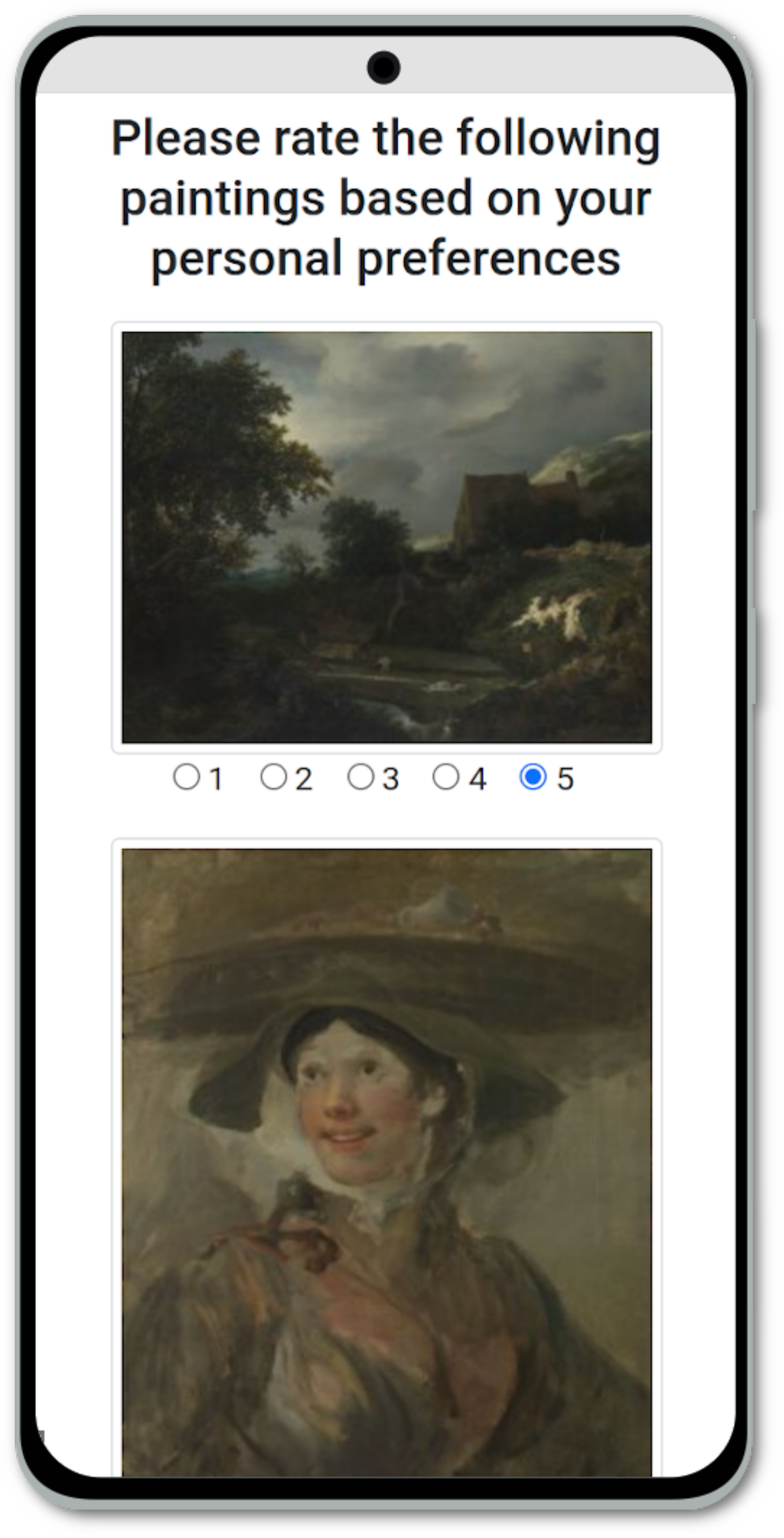} \hfil
    \includegraphics[height=\h]{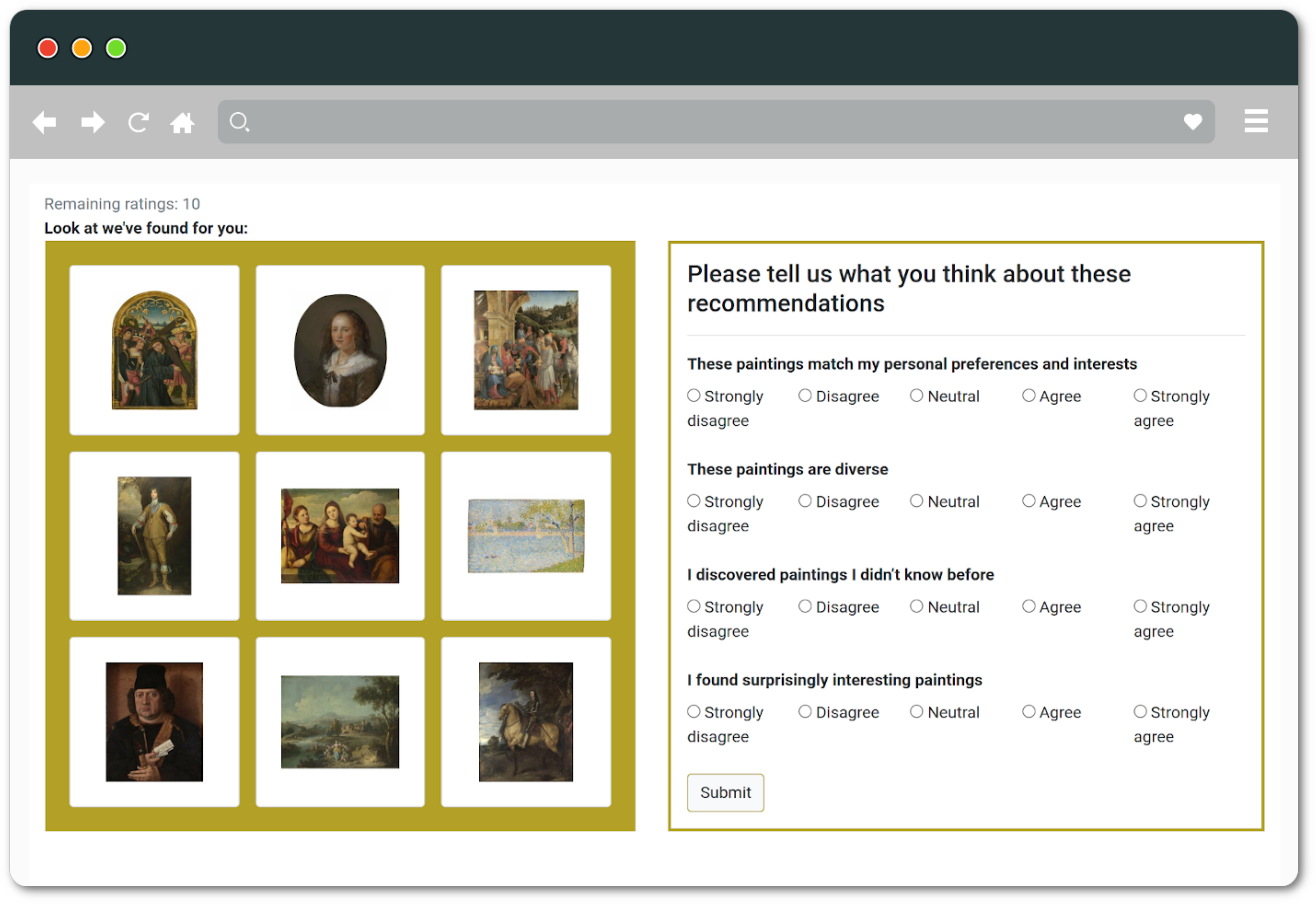}

    \caption{
        Screenshots of our web application for evaluation.
        Left: elicitation screen in mobile mode.
        Right: Recommendation evaluation screen in laptop mode.
    }
    \label{fig:webapp}
\end{figure*}

We created a web application that first collected preference elicitation ratings from users
and then showed a set of VA recommendations based on their elicited preferences.
As shown in \autoref{fig:webapp}, participants were provided with one set of recommendations from each VA RecSys engine at a time.

Since participants could use any device (desktop computer, laptop, or mobile) to complete the study,
we used a responsive design; see \autoref{fig:webapp}.
By clicking or tapping on any image, in both the elicitation and rating screens,
a modal window displays an enlarged version of the image.

\subsection{Participants}

As described in the next section,
we first conducted a small-scale study ($N=11$) with museum visitors,
to gather insights from real-world usage of our application,
and then we conducted a large-scale study ($N=100$) with a carefully selected pool of crowdworkers.

\subsection{Design}

Participants were exposed to all VA engines exactly once (within-subjects design)
and rated the provided recommendations in a 5-point Likert scale.
Our dependent variables are widely accepted proxies of recommendation quality~\cite{pu2011user}:
\begin{description}
    \item[Accuracy:] The paintings match my personal preferences and interests.
    \item[Diversity:] The paintings are diverse.
    \item[Novelty:] I discovered paintings I did not know before.
    \item[Serendipity:] I found surprisingly interesting paintings.
\end{description}

\subsection{Procedure}

Participants accessed our web application and entered their demographics information (age, gender) on a welcome screen.
There, they were informed about the purpose of the study and the data collection policy.
They also indicated their visiting style, for which we adopted the framework proposed by Veron et al.~\cite{veron1989ethnographie}
to classify museum visitors into four visiting style metaphors~\cite{kuflik12profiles},
related to the time they spend during visits:
\begin{description}
    \item[Ant:] I spend a long time observing all exhibits
        and move close to the walls and the exhibits avoiding empty space.
    \item[Fish:] I walk mostly through empty space making just a few stops
        and see most of the exhibits but for a short time.
    \item[Grasshopper:] I see only exhibits I am interested in.
        I walk through empty space and stay for a long time only in front of selected exhibits.
    \item[Butterfly:] I frequently change the direction of my tour, usually avoiding empty space.
        I see almost all exhibits, but time varies between exhibits.
\end{description}

Then, participants advanced to the preference elicitation screen,
where they were shown one painting at random from each of the nine curated story groups.
They rated each painting in a 5-point numerical scale (5 is better, i.e. the user likes the painting the most).
Finally, users advanced to the RecSys assessment screen,
where they were shown a set of nine painting recommendations drawn from each VA RecSys engine.
Note that each user initially rated nine paintings (one from each story group)
but recommendations may come from only one or a few story groups, depending on their elicited preferences.

\subsection{Museum study}

We physically advertised our call for participants in the museum Centre Pompidou-Metz, France
with a flyer that had a QR code for people to scan in order to access the study.
A small sample of $N=11$ participants (6 female, 5 male) aged 36 years (SD=20.8)
voluntarily took part in the study.
The study took 4.7\,min on average to complete (SD=4.3).

\begin{figure*}[!ht]
    \centering
    \def\w{0.24\linewidth}
    \subfloat[Accuracy ratings]{\includegraphics[width=\w]{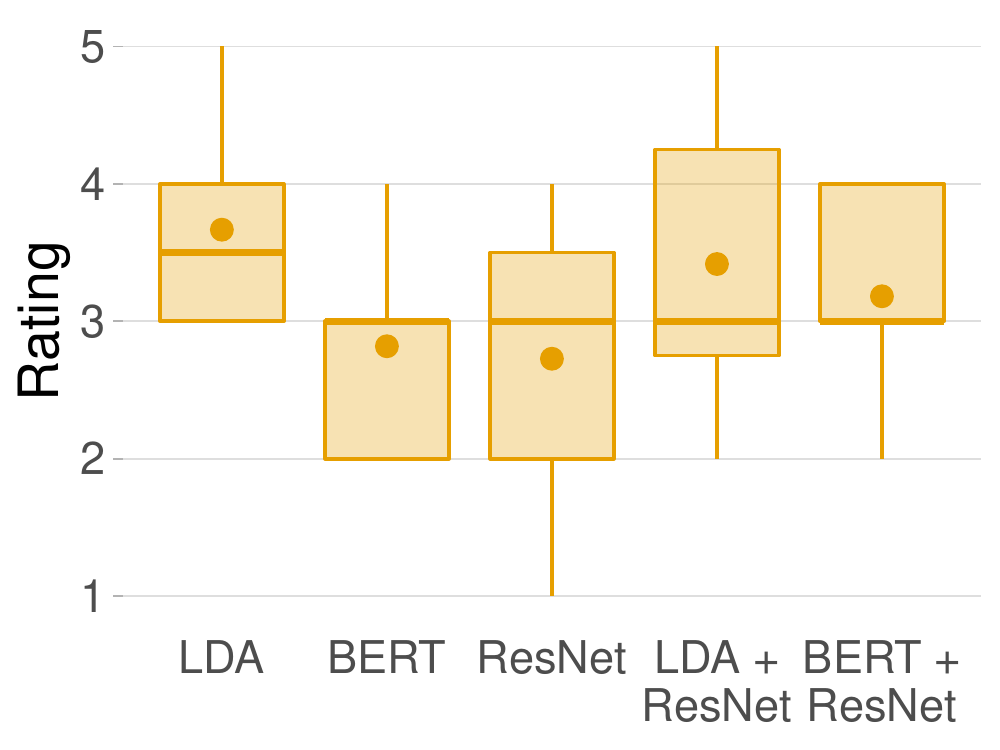}} \hfill
    \subfloat[Diversity ratings]{\includegraphics[width=\w]{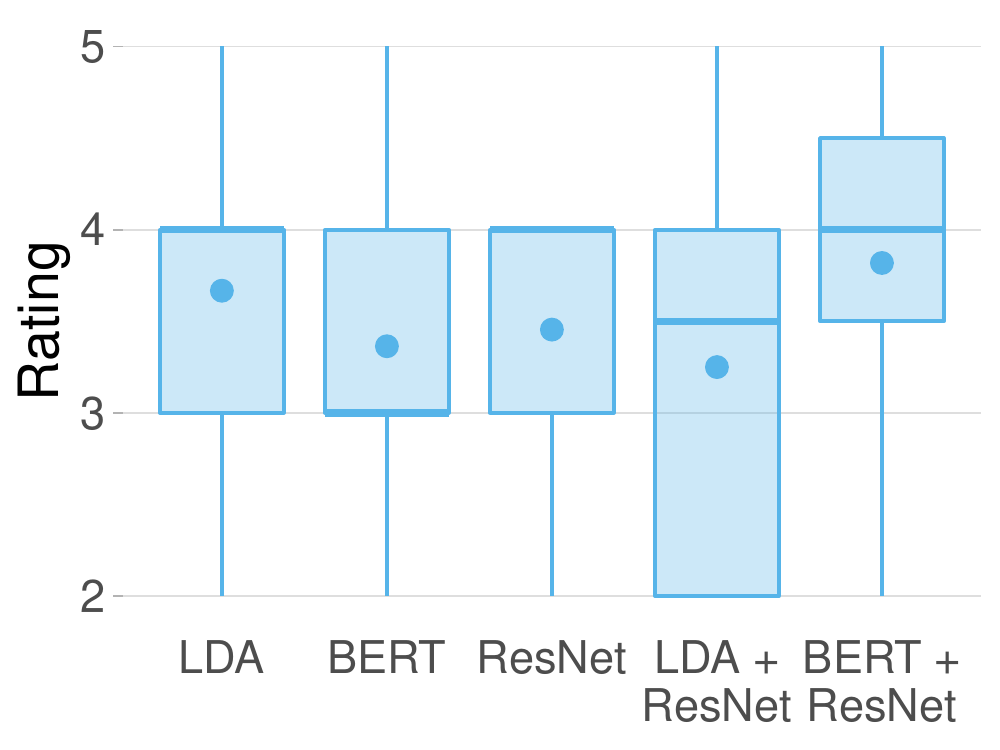}} \hfill
    \subfloat[Novelty ratings]{\includegraphics[width=\w]{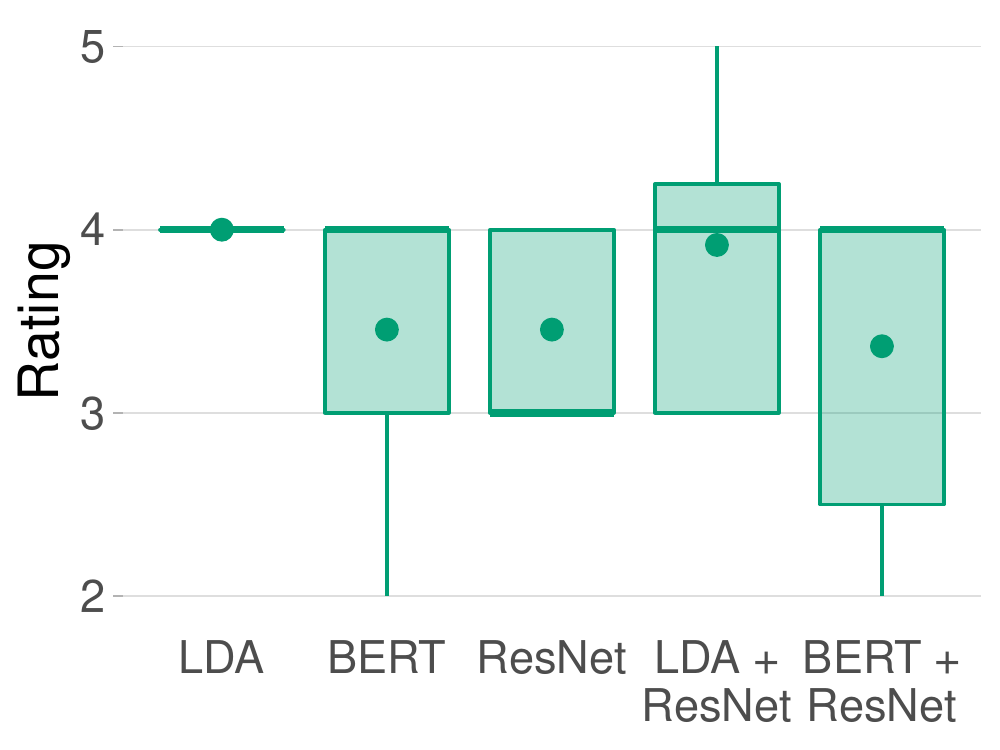}} \hfill
    \subfloat[Serendipity ratings]{\includegraphics[width=\w]{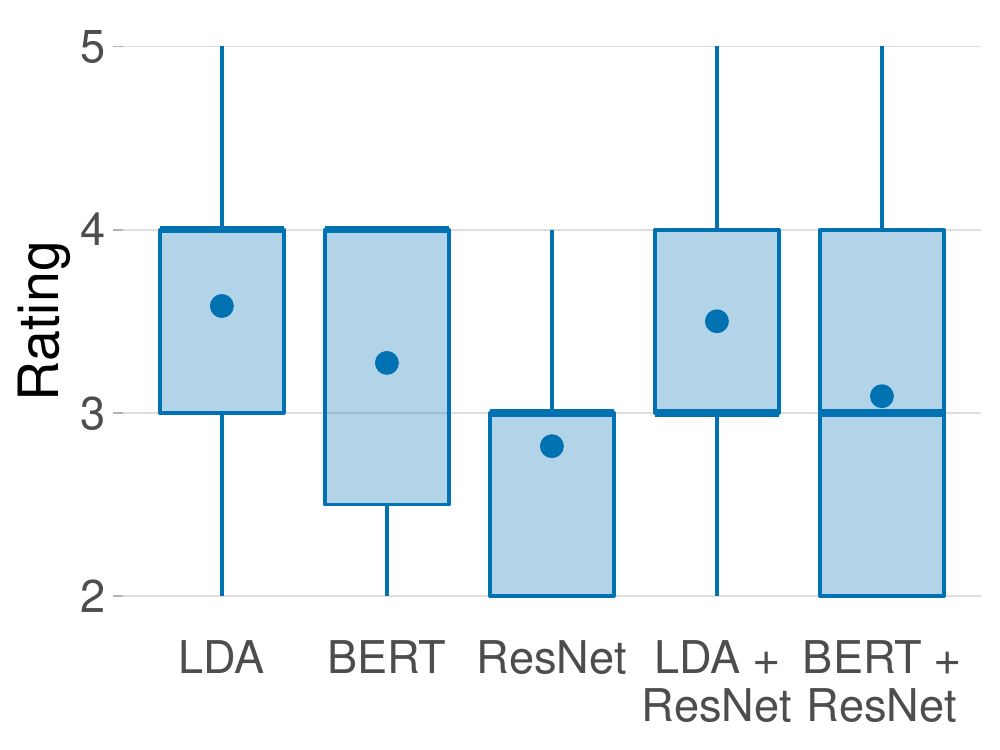}}

    \caption{Distribution of ratings from museum users. Dots denote mean values.}
    \label{fig:museum-ratings-overall}
\end{figure*}

\begin{figure*}[!ht]
    \centering
    \def\w{0.48\linewidth}
    \subfloat[Accuracy ratings]{\includegraphics[width=\w]{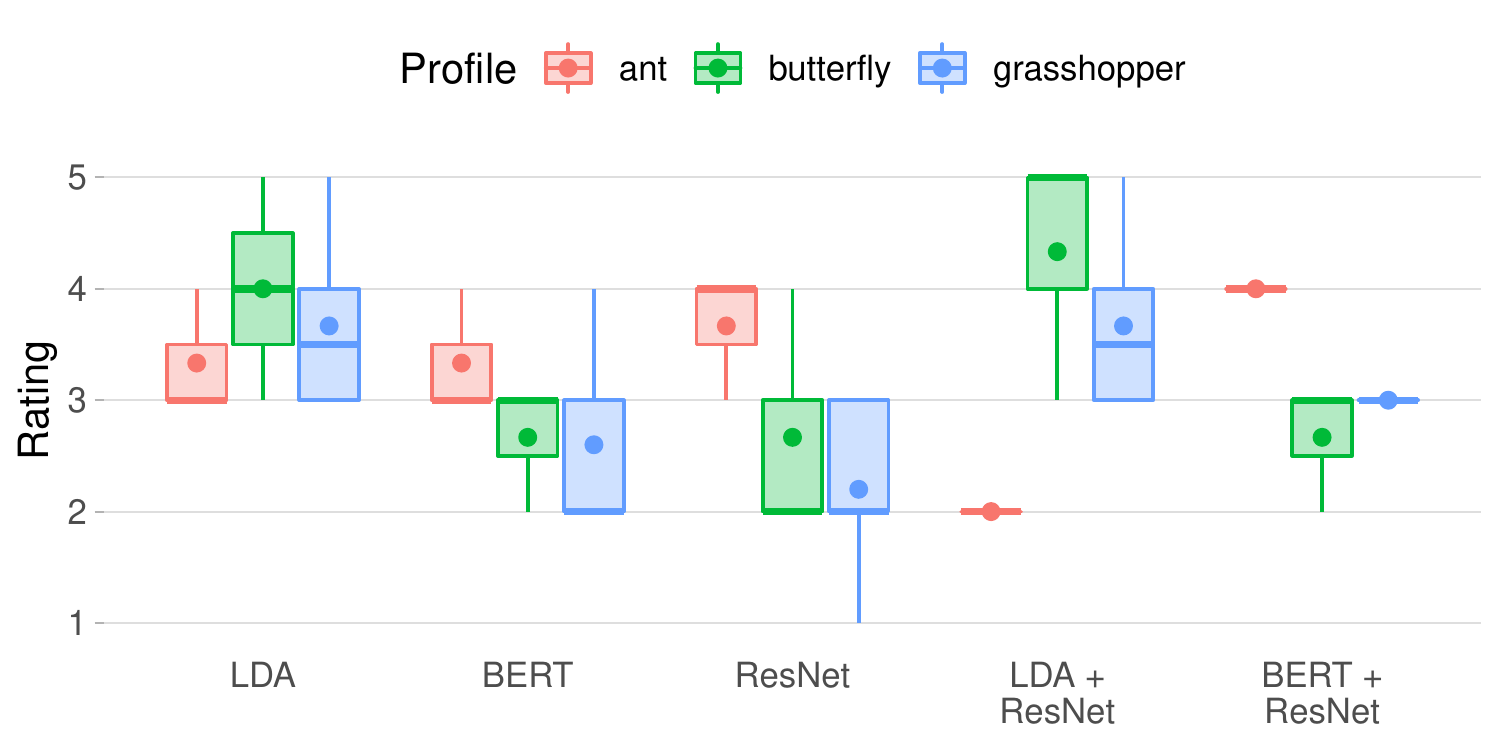}} \hfil
    \subfloat[Diversity ratings]{\includegraphics[width=\w]{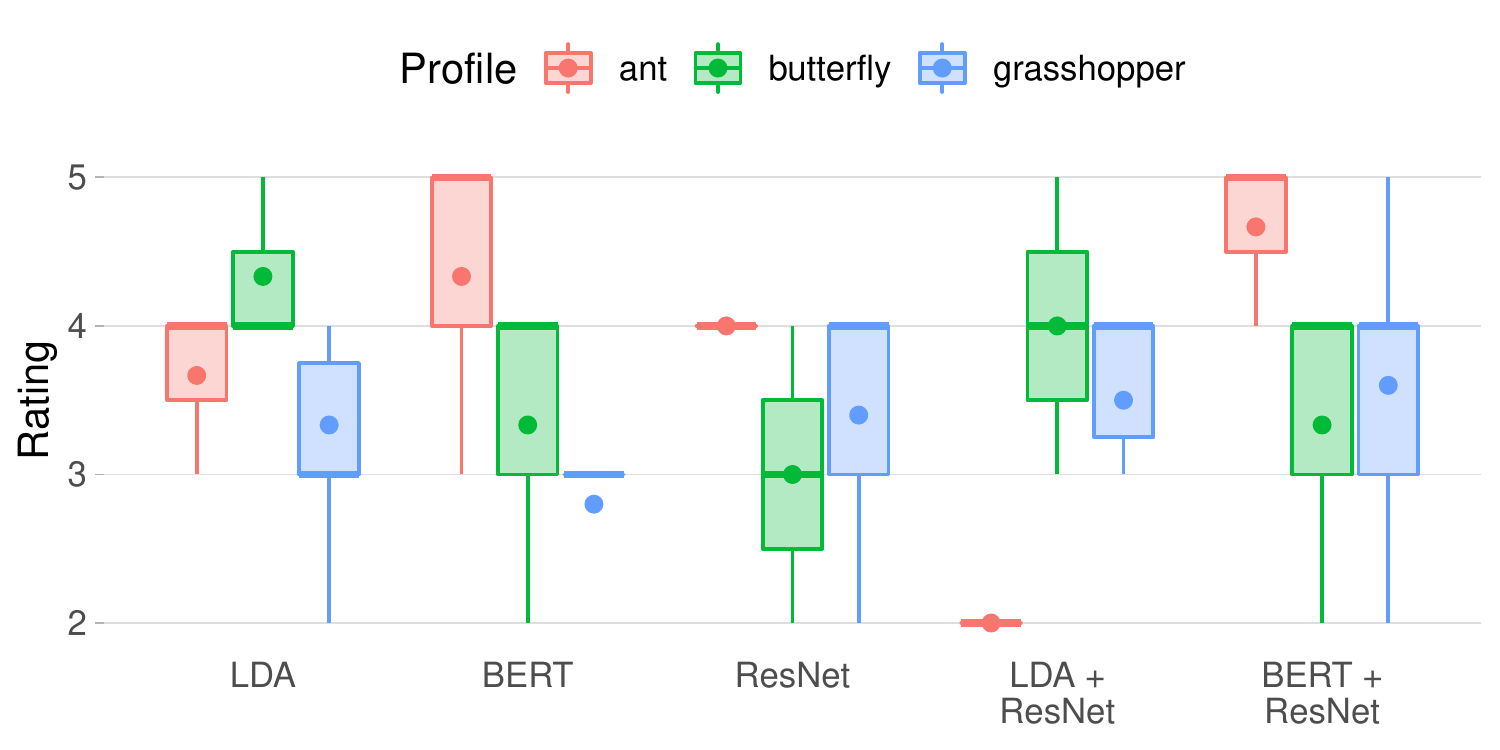}} \\
    \subfloat[Novelty ratings]{\includegraphics[width=\w]{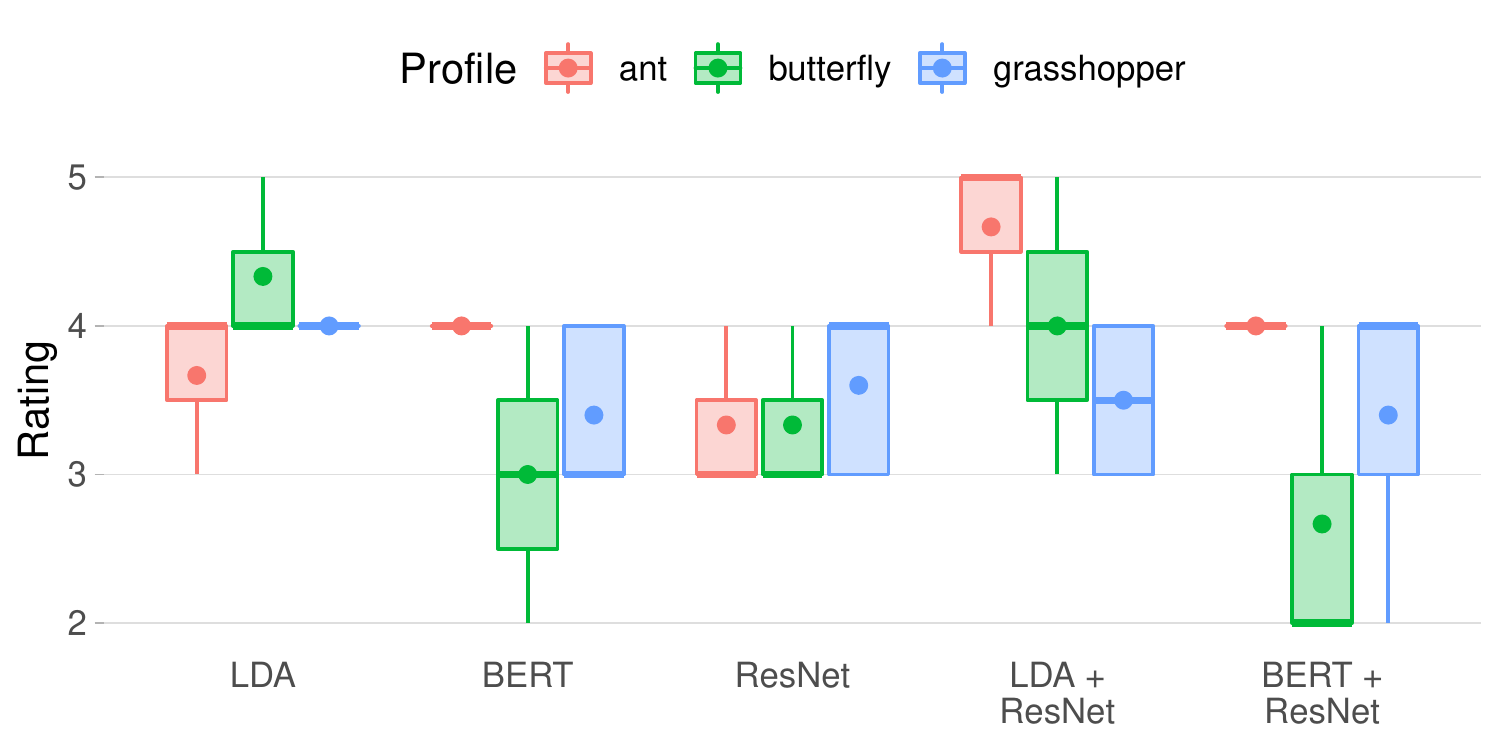}} \hfil
    \subfloat[Serendipity ratings]{\includegraphics[width=\w]{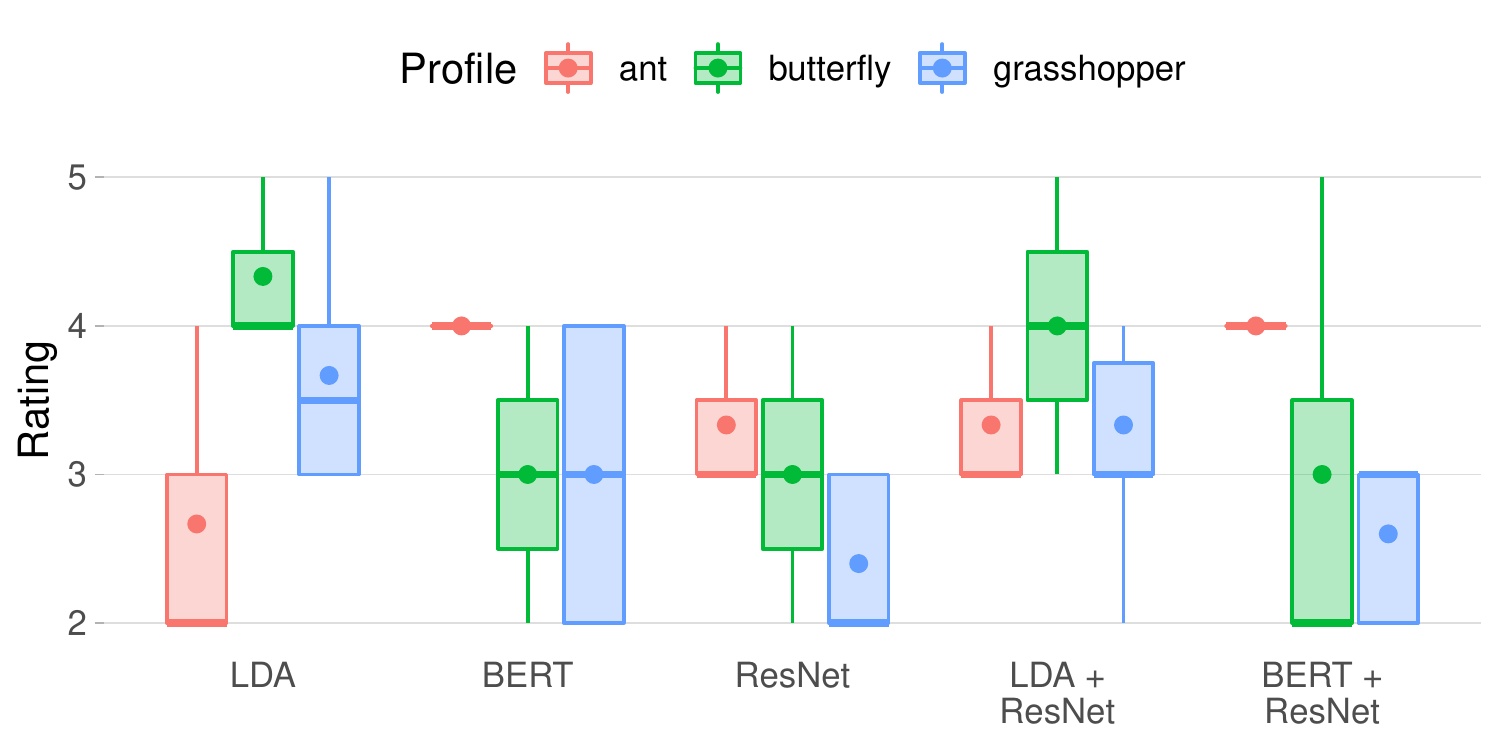}}

    \caption{Distribution of ratings from museum users, segregated by visiting profiles. Dots denote mean values.}
    \label{fig:museum-ratings-groups}
\end{figure*}

\autoref{fig:museum-ratings-overall} shows the distributions of user ratings for each of the dependent variables considered.
\autoref{fig:museum-ratings-groups} segregates the results by the different visiting profiles.
We can see that participants perceived each VA engine differently for each of the evaluation metrics considered.
For example, LDA was rated the highest in terms of Accuracy and Novelty,
whereas the fusion of BERT+ResNet was rated higher in terms of Diversity.
Interestingly, ResNet was rated the lowest in terms of Serendipity.

We investigated whether there is any difference between any of the five RecSys engines,
for which we use a linear mixed-effects (LME) model where each dependent variable
is explained by each VA RecSys engine.
The visiting profile is considered an interaction effect (model covariate)
and participants are considered random effects.
An LME model is appropriate here because the dependent variables are discrete and have a natural order.
In addition, LME models are quite robust to violations of several distributional assumptions~\cite{Schielzeth20}.

We fit the LME models (one per dependent variable) and compute the estimated marginal means for specified factors.
We then run pairwise comparisons (also known as \emph{contrasts} in LME parlance)
with Bonferroni-Holm correction to guard against multiple comparisons.\footnote{
The Bonferroni-Holm correction method sorts $p$-values from lowest to highest
and compares them to nominal alpha levels of $\frac{\alpha}{m}$ to $\alpha$.
Then, it finds the index $k$ that identifies the first p-value that is not low enough to validate rejection of the null hypothesis.
}
We observed that LDA was significantly preferred over BERT ($p=.028, r=0.449$)
and ResNet ($p=.028, r=0.459$) engines in terms of Accuracy.
LDA was preferred over ResNet ($p=.048, r=0.459$)
as well as over the fusion of BERT+ResNet ($p=.046, r=0.409$) in terms of Novelty.
The LDA+ResNet engine outperformed BERT ($p=.048, r=0.379$) and ResNet ($p=.048, r=0.404$)
as well the fusion of BERT+ResNet ($p=.046, r=0.439$) in terms of Novelty.
All other comparisons were not found to be statistically significant.
However, effect sizes ($r$, analogous to Cohen's $d$)
suggest a moderate importance of the differences between RecSys engines in practice.
For example, LDA was preferred over BERT in terms of Novelty ($r=0.347, p=.060$)
and the fusion of LDA+ResNet was preferred over BERT+ResNet in terms of Diversity ($r=0.338, p=.357$).
ResNet was less preferred than LDA or LDA+ResNet in terms of Serendipity ($r=0.335, p=.223$).

If we take closer look at the results per visiting profiles (\autoref{fig:museum-ratings-groups}),
we can observe that Ant users prefer BERT topics over LDA topics, and this is also reflected in the fused rankings.
For example, in terms of Accuracy, Diversity and Serendipity,
Ant users ranked BERT-based recommendations higher than Butterfly and Grasshopper users.
On the other hand, Grasshopper users did not like BERT-based recommendations overall.
Instead, in terms of Novelty, the fusion of LDA+ResNet was preferred over BERT+ResNet.
We observed a statistically significant correlation between visitor profiles and ratings
in terms of Diversity ($\rho=0.26, p < .01$) and Serendipity ($\rho=0.3, p < .001$).
This can potentially be an indication that the visiting style of the user,
to a certain extent, reflects their preferences towards art content.
Hence, it could be leveraged to parameterise different aspects of RecSys
(e.g, Diversity, Novelty, etc.) in future work.

\subsection{Crowdsourcing study}

We recruited a large sample of $N=100$ participants via the Prolific crowdsourcing platform.\footnote{\url{https://www.prolific.co/}}
We enforced the following screening criteria for any participant to be eligible:
\begin{itemize}
\item The primary language is English.
\item Art is listed among their interests/hobbies.
\item Minimum approval rate of 99\% in previous crowdsourcing studies in the platform.
\item Registration date before January 2022.
\end{itemize}

Our recruited participants (75 female, 25 male) were aged 39.7 years (SD=14.1) and could complete the study only once.
Most of them had UK nationality (59\%) or were living in the UK (64\%).
The study took 5.7\,min on average to complete (SD=2.3) and participants were paid an equivalent hourly wage of \$10/h.

\begin{figure*}[!ht]
    \centering
    \def\w{0.24\linewidth}
    \subfloat[Accuracy ratings]{\includegraphics[width=\w]{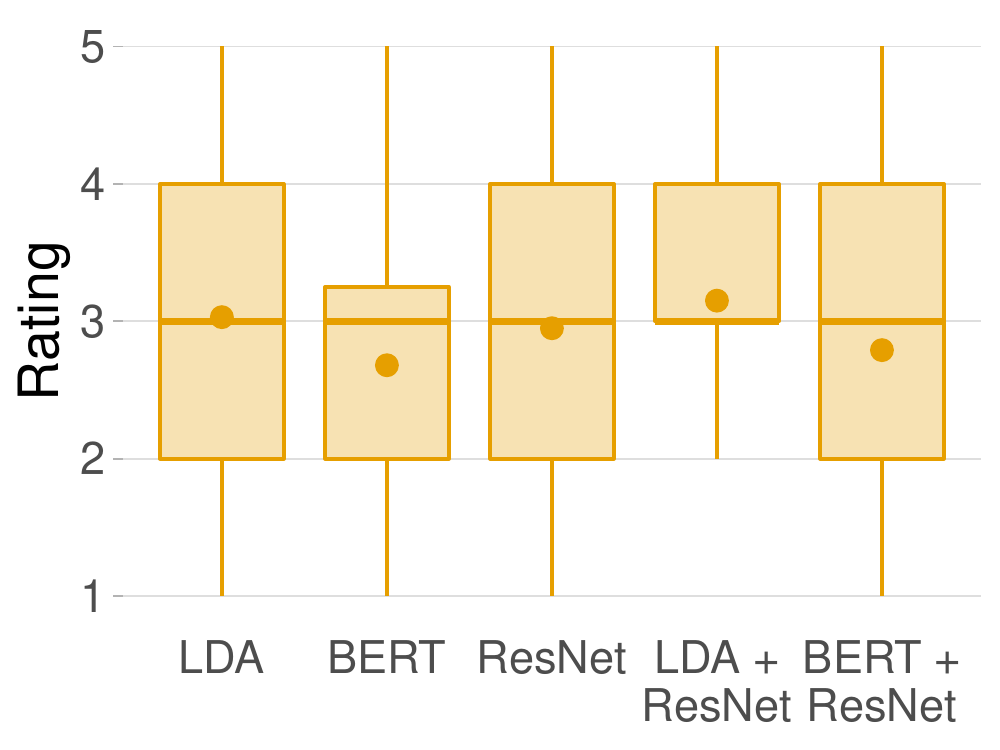}} \hfill
    \subfloat[Diversity ratings]{\includegraphics[width=\w]{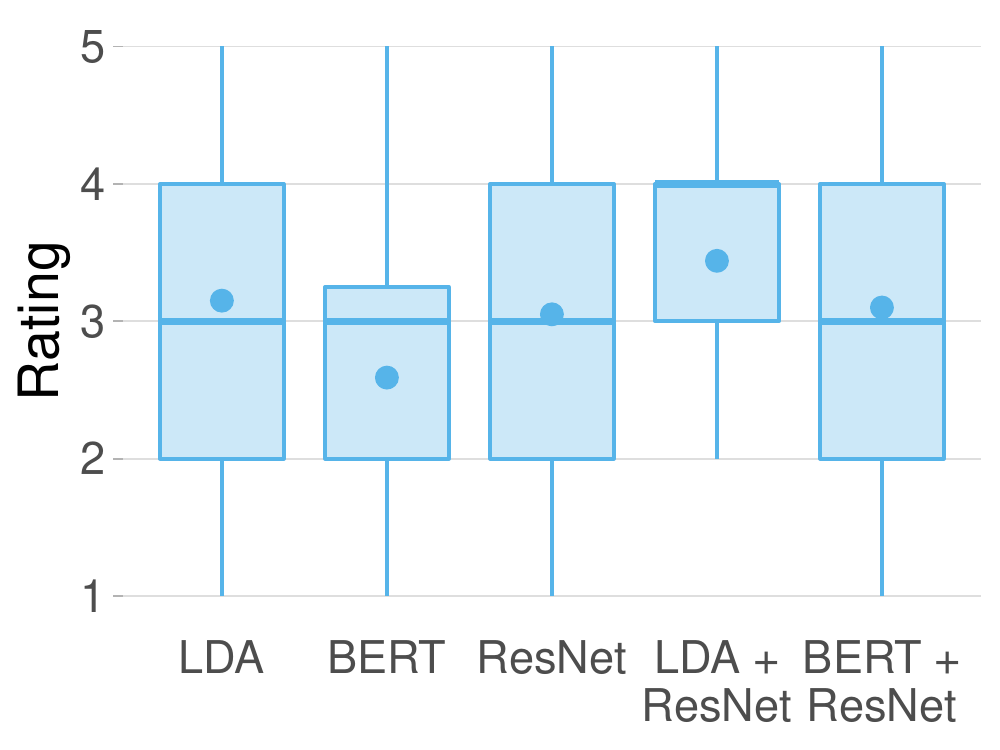}} \hfill
    \subfloat[Novelty ratings]{\includegraphics[width=\w]{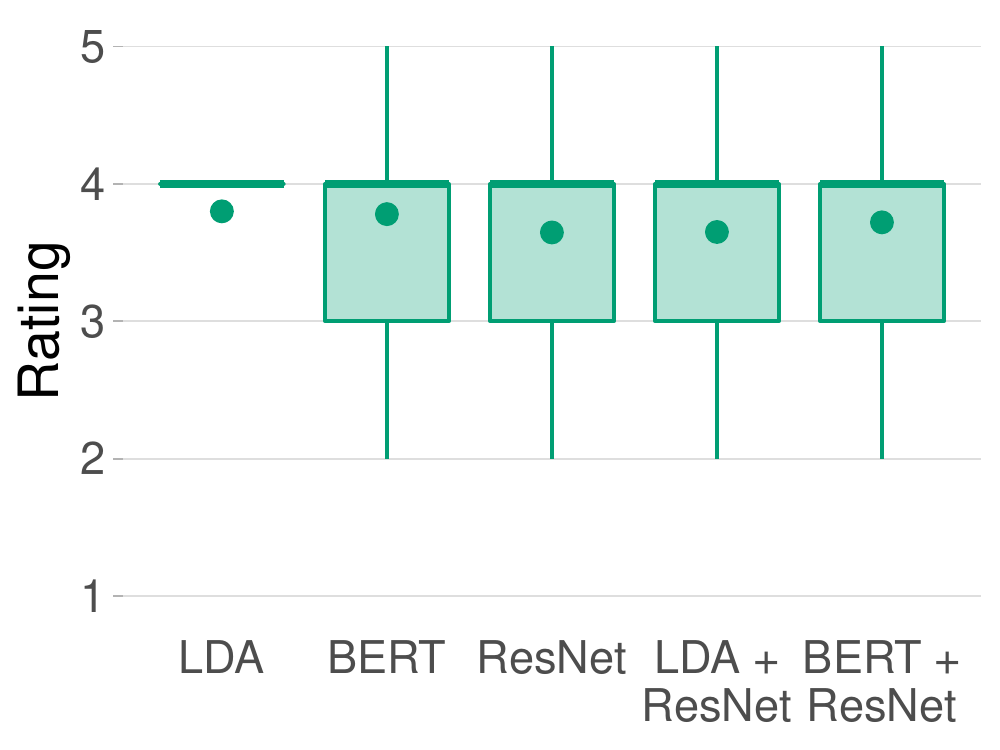}} \hfill
    \subfloat[Serendipity ratings]{\includegraphics[width=\w]{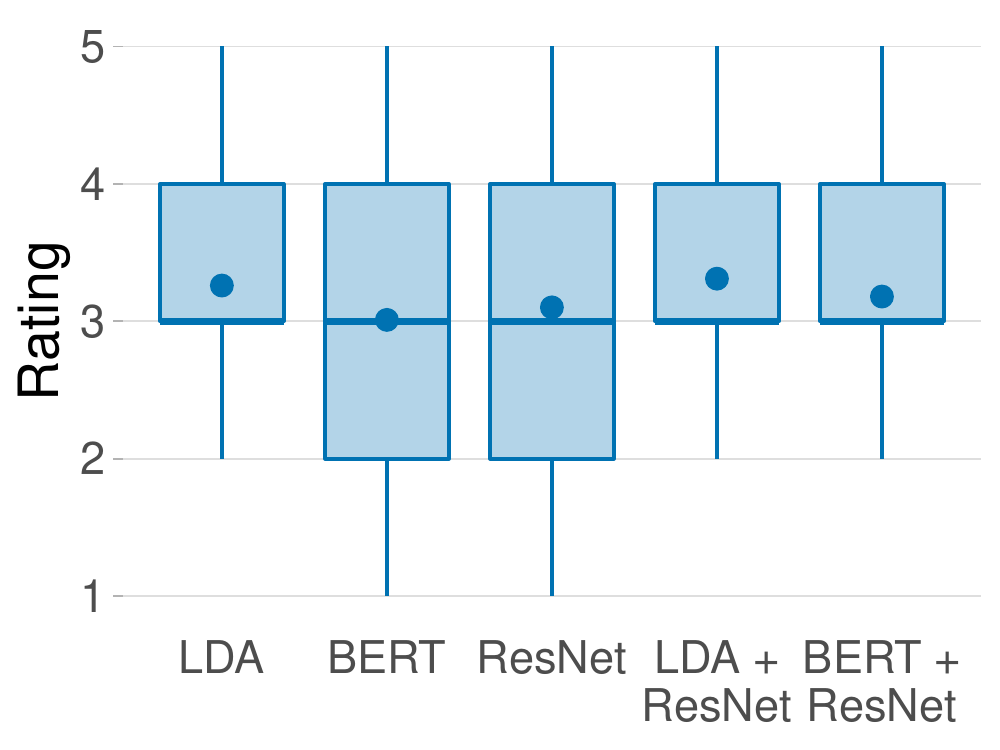}}

    \caption{Distribution of ratings from crowdsourcing users. Dots denote mean values.}
    \label{fig:prolific-ratings-overall}
\end{figure*}

\begin{figure*}[!ht]
    \centering
    \def\w{0.48\linewidth}
    \subfloat[Accuracy ratings]{\includegraphics[width=\w]{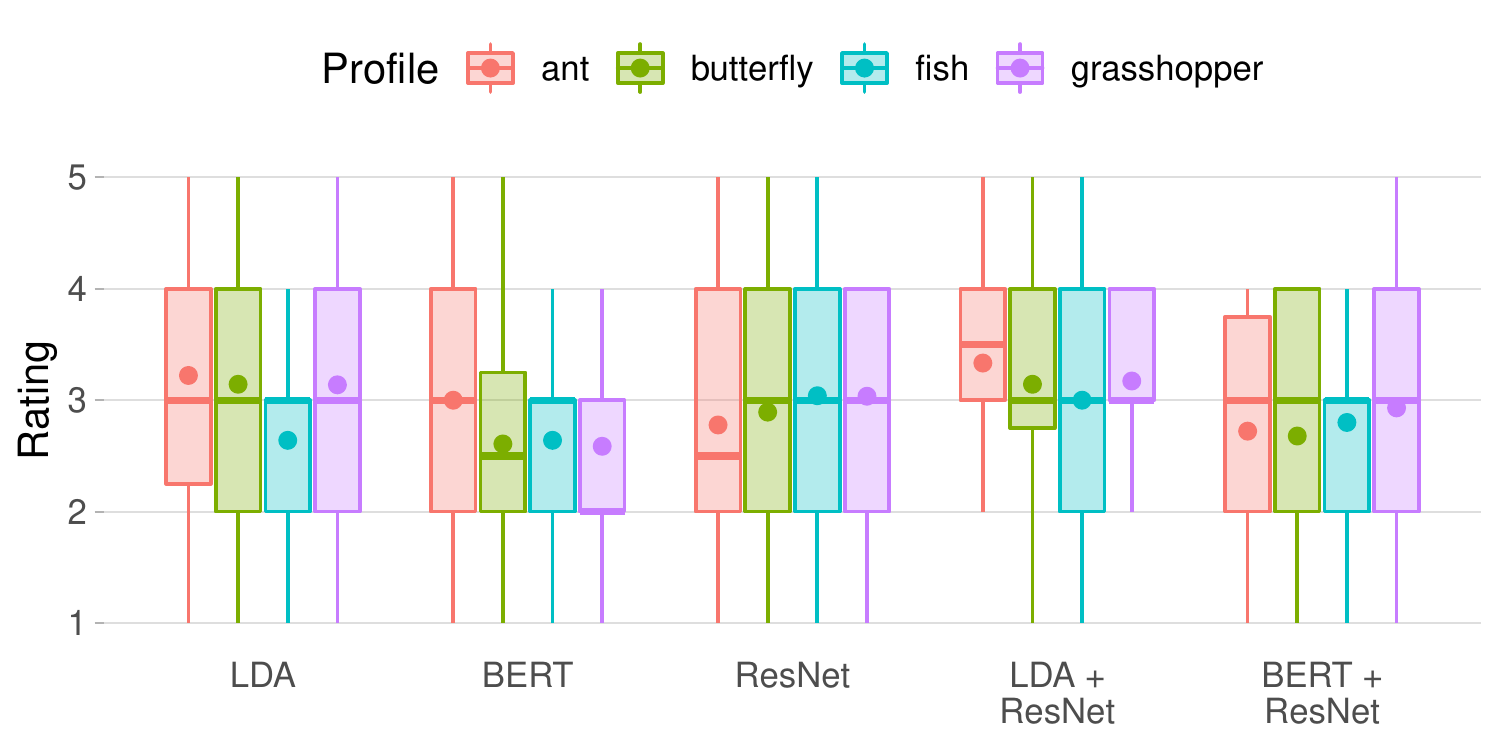}} \hfil
    \subfloat[Diversity ratings]{\includegraphics[width=\w]{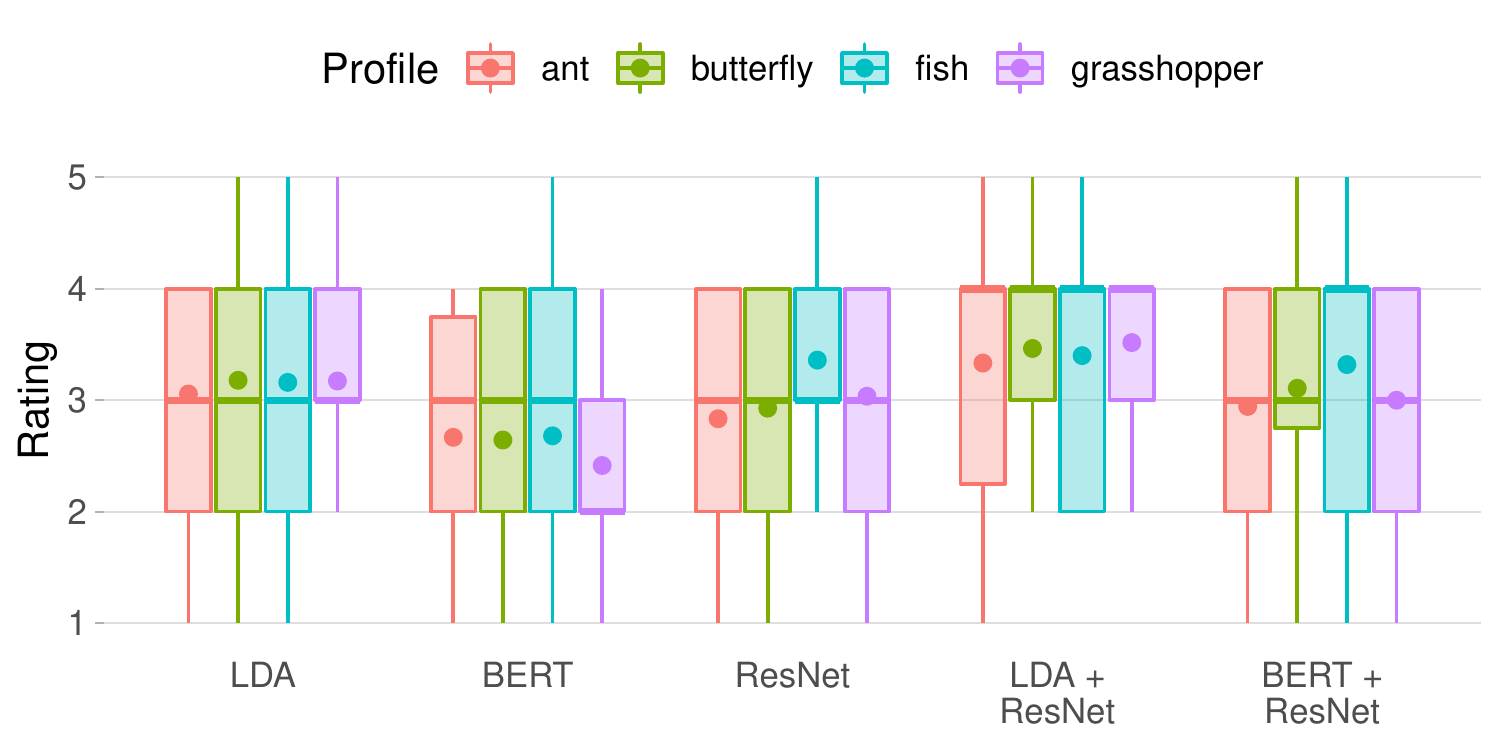}} \\
    \subfloat[Novelty ratings]{\includegraphics[width=\w]{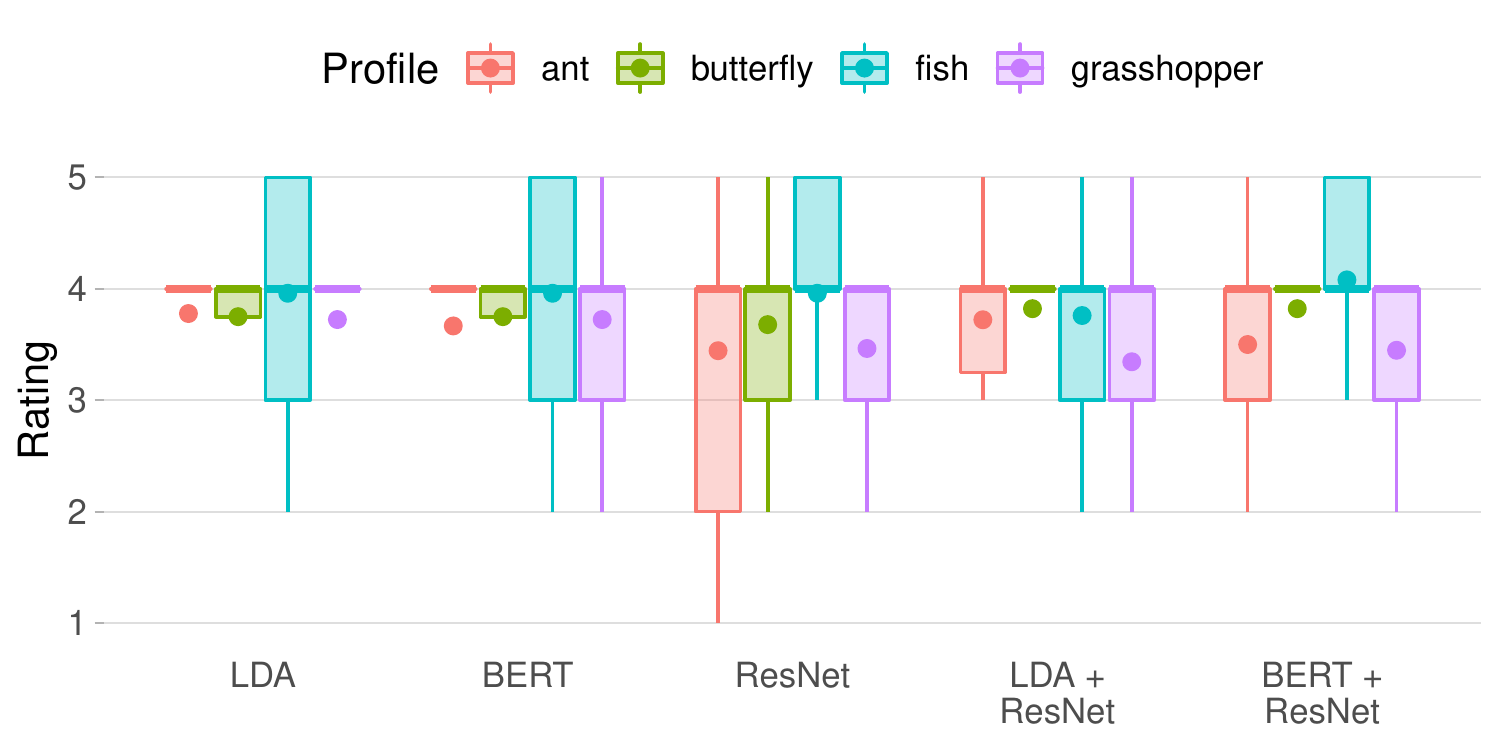}} \hfil
    \subfloat[Serendipity ratings]{\includegraphics[width=\w]{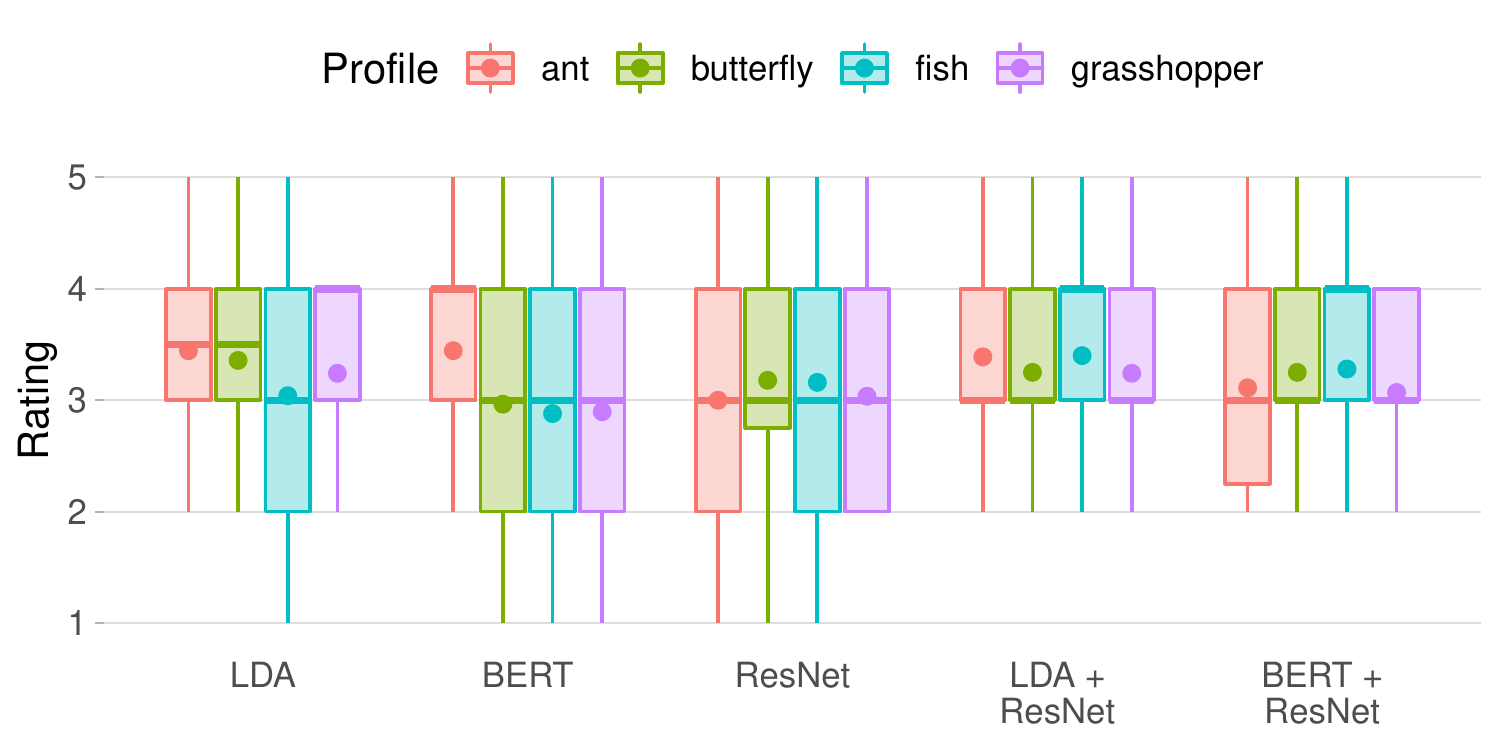}}

    \caption{Distribution of ratings crowdsourcing users, segregated by visiting profiles. Dots denote mean values.}
    \label{fig:prolific-ratings-groups}
\end{figure*}

\autoref{fig:prolific-ratings-overall} shows the distributions of user ratings for each of the dependent variables considered.
\autoref{fig:prolific-ratings-groups} segregates the results by the different visiting profiles.
We can see that, overall, crowdworkers tended to rate the VA RecSys engines slightly higher than museum users.
We observed that the fusion of LDA+ResNet delivered the highest-quality results,
as the ratings received had the narrower inter-quartile difference.
This was systematically so for all the four evaluation metrics considered; see \autoref{fig:prolific-ratings-overall}.

As in the previous study, we fit the LME models and compute the estimated marginal means for specified factors.
We then run pairwise comparisons with Bonferroni-Holm correction to guard against over-testing the data because of the multiple comparisons.
We observed that BERT was significantly less preferred than LDA ($p=.033, r=0.131$)
and LDA+ResNet ($p=.003, r=0.18$) in terms of Accuracy.
BERT+ResNet was outperformed by LDA+ResNet in terms of Accuracy ($p=.014, r=0.151$).
In terms of Diversity, BERT was rated significantly lower than any other approach ($p<.001, 0.196 < r < 0.36$)
and the fusion of LDA+ResNet outperformed BERT+ResNet ($p<.01, r=0.154$)
as well as the individual LDA ($p=.013, r=0.132$) and ResNet ($p<.001, r=0.181$) engines.
All other comparisons were not found the be statistically significant.
We can conclude therefore that the fusion of text and image features is the most beneficial approach
to deliver more adequate recommendations to the user.

In this crowdsourcing study we did not observe strong correlations between user profiles and ratings.
However, a few interesting observations can be made.
For example, from the results per visiting profiles (\autoref{fig:prolific-ratings-groups})
we can see that Fish users did not like BERT-based recommendations, which was also reflected in the fused ranking BERT+ResNet.
In terms of Diversity, Grasshopper users prefer LDA over BERT.
This can be attributed to the larger topic size in LDA (10 topics) compared to BERT (4 topics).
Hence, we hypothesise that users who preferred LDA are most likely interested in diverse VA content,
especially it we take into account that Grasshopper profiles have a clear expectation of what to find in a museum.
In terms of Novelty, Butterfly users showed more agreements in their rankings,
as the interquartile range is much smaller as compared to the other visiting profiles.
Finally we observed that Fish users tended to provide higher ratings than the other user profiles,
especially for ResNet and BERT+ResNet recommendations.
As discussed in the previous section, these observations could potentially inform novel ways of operationalising different aspects of RecSys in future work.

\subsection{Ranking overlap analysis}

We conducted an additional analysis that checked whether the users were receiving truly personalized recommendations.
Otherwise, our VA RecSys engines would have been recommending the same contents to every user.
To account for this, we compute the Intersection over Union (IoU) and Rank-Biased Overlap (RBO),
which are widely used measures in information retrieval~\cite{10.1145/1852102.1852106}. RBO and IoU were calculated in a pairwise manner among all users exposed to the same engine and averaged.
\autoref{tab:ranking_overlap} presents the results of this analysis.
As shown in the table, there is no substantial overlap in the rankings produced by each engine.
This analysis indicates that each user indeed was shown a personalized set of recommendations.

\begin{table*}[!h]
\centering
\caption{Ranking overlap results, showing Mean $\pm$ SD of IoU and RBO measures.}
\label{tab:ranking_overlap}
\begin{tabular}{ll *6c}
\toprule
& &
  \textbf{LDA} &
  \textbf{BERT} &
  \textbf{ResNet} &
  \textbf{LDA+ResNet} &
  \textbf{BERT+ResNet} &
  \textbf{All} \\
\midrule
Crowdsourcing &
  IoU &
  0.09 $\pm$ 0.15 &
  0.09 $\pm$ 0.15 &
  0.09 $\pm$ 0.15 &
  0.09 $\pm$ 0.15 &
  0.09 $\pm$ 0.15 &
  0.07 $\pm$ 0.11 \\
study &
  RBO &
  0.10 $\pm$ 0.16 &
  0.10 $\pm$ 0.16 &
  0.10 $\pm$ 0.17 &
  0.09 $\pm$ 0.16 &
  0.09 $\pm$ 0.16 &
  0.07 $\pm$ 0.12 \\
   \midrule
Museum &
  IoU &
  0.27 $\pm$ 0.26 &
  0.33 $\pm$ 0.26 &
  0.32 $\pm$ 0.26 &
  0.27 $\pm$ 0.27 &
  0.33 $\pm$ 0.26 &
  0.11 $\pm$ 0.16 \\
study &
  RBO &
  0.26 $\pm$ 0.27 &
  0.31 $\pm$ 0.27 &
  0.31 $\pm$ 0.27 &
  0.26 $\pm$ 0.27 &
  0.31 $\pm$ 0.26 &
  0.09 $\pm$ 0.16 \\
\bottomrule
\end{tabular}%
\end{table*}

\section{Discussion}

From a conceptual point of view, this paper has advanced our understanding of how users perceive and evaluate VA RecSys.
In recent years, the research community has shifted to include a wider range of ``beyond accuracy'' objectives~\cite{Kaminskas16},
such as the user-centric dependent variable we have used in our studies,
however the field of VA personalization has remained largely unexplored in this regard.
We have found that text-only and vision-only RecSys compare similarly in terms of recommendation quality,
but the fusion of these two approaches delivers the best results.
We also have observed that different visiting style profiles may benefit differently from each type of recommendations,
although fusion-based recommendations are systematically preferred overall.

Previous work suggested that visual features are preferred over textual features
when it comes to delivering high-quality VA recommendations to the users~\cite{messina2017exploring, messina2019content, messina2020curatornet}.
However, our experiments have demonstrated that they provide similar results.
This was so for the small-scale and the large-scale study.
Therefore, we reject \textbf{H1} and conclude that visual features perform no better than textual features.
This is somehow understandable, since each type of latent representation
provides a different understanding about the paintings. Furthermore, the improved performance observed in the fusion approaches indicates that both visual and textual features complement each other to efficiently capture the elements of VA RecSys,
which leads us to validate \textbf{H2}.
In the following, we provide a critical and in-depth discussion about our results
and what they imply for the HCI community.

\subsection{Visual similarity does not entail semantic similarity (and vice versa)}
\label{subsec:discuss_one}

Nowadays, with the recent advances in computer vision, capturing visual similarity of images is relatively an effortless task. Hence, finding visually similar paintings to what users previously saw or expressed interest seems straightforward. However, as discussed in \autoref{sec:related-work}, understanding users' perception of artwork is an extremely challenging task due to the complexity of concepts embedded within the artworks as well as the reflections they may trigger on users. Contrary to most prominent work in VA RecSys that leveraged only visual features to derive recommendations, we explored textual features as well as hybrid approaches combining the learned text-based and image-based features. Interestingly, our work provides compelling evidence that visual similarity does not necessarily entail semantic relatedness.

\begin{figure*}[!ht]
\centering
\includegraphics[width=\textwidth]{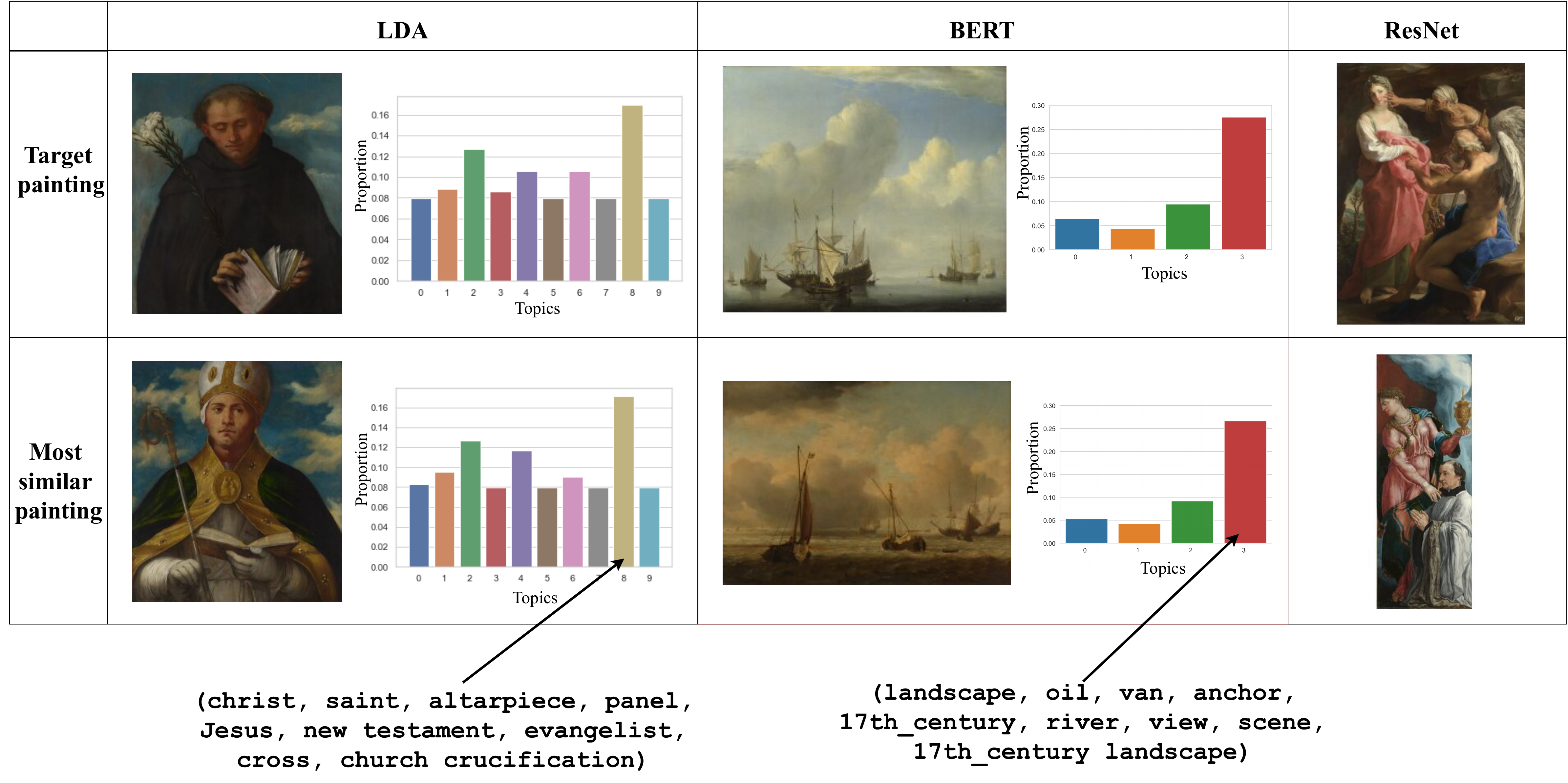}
\caption{Examples of target paintings (top) and most similar paintings (bottom) according to LDA, BERT and ResNet.}
\label{fig:sematics}
\end{figure*}

In \autoref{fig:sematics} we illustrate this phenomenon with examples. We show a target painting (top) and its most similar painting (bottom) according to the three VA RecSys engines. For LDA and BERT we additionally show the paintings' topic distributions and their descriptions.  For LDA (first column) we can see that paintings have very similar topic distribution and topic 8 stands out. This implies that words in topic 8 are more likely to be found in the paintings descriptions than the words from the other topics. Actually topic 8 is very well defined as there is high coherence between the words. In fact, topic 8 can be described as a ``Christian'' topic of the collection, since many of the words in this topic are usually found in christian corpora such as biblical texts. When looking at the paintings, there are many references to Christianity, therefore we can assume that their descriptions contain vocabulary that refers to a religious context. The ground-truth from the National Gallery documentation also supports this claim, as both paintings are from the panels of the high altarpiece of the church of Sant'Alessandro Brescia, painted by Girolamo Romanino in the 16th century. Then, the target painting\footnote{https://www.nationalgallery.org.uk/paintings/girolamo-romanino-saint-filippo-benizzi} shows Saint Filippo Benizzi, who was the fifth general of the Servites, the order to whom the church belonged. The most similar painting according to LDA\footnote{https://www.nationalgallery.org.uk/paintings/girolamo-romanino-saint-gaudioso} is a portrait of Saint Gaudioso, who was the bishop of Brescia in the 5th century, and was buried in the church.

For BERT (second column), the target painting is ``Calm: A Dutch Ship coming to Anchor and Another under Sail''\footnote{https://www.nationalgallery.org.uk/paintings/willem-van-de-velde-a-dutch-ship-coming-to-anchor} by Willem van de Velde, and the most similar one according BERT is ``Dutch Ships and Small Vessels Offshore in a Breeze''\footnote{https://www.nationalgallery.org.uk/paintings/willem-van-de-velde-dutch-ships-and-small-vessels-offshore-in-a-breeze} by the same artist. When we look at how BERT represents these two paintings, we can observe that they have very similar topic distributions. Particularly topic 3 is very prominent in both paintings. Taking a closer look at the topic descriptions, we can understand that BERT created a coherent representation. Observing the actual images of the paintings, we can also tell that the paintings are visually very similar. Overall, both examples of LDA and BERT demonstrate that similarities of visual features can be captured from semantic similarities of textual features. However, our analysis on ResNet shows that the inverse is not necessarily true.

The last column in \autoref{fig:sematics} illustrates a sample target painting and its most similar painting according to ResNet.
The target is a painting from the 18th century titled ``Time orders Old Age to destroy Beauty'' \footnote{https://www.nationalgallery.org.uk/paintings/pompeo-girolamo-batoni-time-orders-old-age-to-destroy-beauty} by Pompeo Girolamo Batoni. In this case, the most similar painting is from 16th century titled ``The Donor and Saint Mary Magdalene''\footnote{https://www.nationalgallery.org.uk/paintings/marten-van-heemskerck-the-donor-and-saint-mary-magdalene} by Marten van Heemskerck. Looking at the two paintings, without further context, one can easily tell that ResNet manages to capture visual features such as colors, edges, and corners among the paintings. However, the two paintings are not very semantically related. The target painting depicts ``time'' by the winged figure holding an hourglass, ordering his companion Old Age to disfigure the face of a young woman, the personification of Beauty. The National Gallery documentation states: \emph{With this painting, Batoni intends to  encourage considering the brevity of youth and the inevitable passing of time}. On the other hand, \emph{``The Donor'' depicts a statuesque Mary Magdalene, one of Christ's followers, resting her fingers on the shoulder of a kneeling donor, and with the other hand she is nonchalantly lifting a large golden vessel. This is the pot containing the precious ointment with which she anointed Christ's feet (Luke 7:37). In sharp contrast to her colourful opulence, the donor is a serious-looking middle-aged man dressed as a canon}. The National Gallery documentation also mentions that this is one of two shutters from a triptych (a painting made up of three sections), the central part of which is lost.

Given the above discussion, we can deduce that visual similarity does not necessarily entail semantic relatedness. Especially for VA RecSys applications, relying only on visual features can have a negative impact on the quality of recommendations. For example, a user who is not at all interested in religion or Christianity receiving `` The Donor''  as a recommendation just because they liked or previously expressed interest for ``Time orders Old Age to destroy Beauty'' may not be desirable.  Thus, although visual features are important in describing an artwork they alone can not represent the underlying complex semantic relationships.

\subsection{Not all topics are created equal}

In general, the topic distributions learned by a topic model can be used as a semantic representation,
which can be used in several downstream tasks such as document classification, clustering, retrieval, or visualisation~\cite{zhao2021topic}. Particularly, our topic models for VA RecSys have demonstrated the power of exploiting textual data to understand semantic relationships of paintings. This was reflected in the improved performance when combining LDA and BERT with ResNet.  Although LDA and BERT bring in statistical analysis of abstract concepts from the textual data, each technique  has its own uniqueness and relies on different assumptions.

Topic models learn from documents in an unsupervised way and usually measured using a single metric (e.g., topic coherence), which can reflect just one aspect of a model. However, documents are usually associated with rich sets of metadata at both the document and word levels. Overall, evaluating topic models is challenging due to the variety of current frameworks and architectures. It is also evident that quantitative methods are limited in their ability to provide in-depth contextual understanding~\cite{egger2021identifying}. Thus, the interpretation of topic models still relies heavily on human judgment.

\subsection{High-quality recommendations emerge from high-quality latent representations}

The elements of VA recommendation are the latent features of paintings,
therefore is clear that high-quality latent representations must be learned in order to provide high-quality recommendations.
We have shown that each type of RecSys engine (text or image based) is capturing one dimension of the user/painting latent space.
Since paintings are made of visual and textual data, it is beneficial to consider both aspects when generating recommendations to the user.
As mentioned in previous work, the community has been arguing for ignoring textual features~\cite{messina2020curatornet, he2016vbpr, he2016vista} in favor of visual features, however we have shown that doing so will ignore an important dimension of paintings and therefore an important element of visual art recommendation.

\subsection{Knowing the user preferences is key, but it comes at a cost}

Recommender systems require interactions from users to infer personal preferences about new items~\cite{HernandezRubio20}.
It is paramount to know as much information as possible from the users, in particular in the form of ratings,
however we should not burden the users by asking the users to rate every painting they have visited.
Therefore, we must seek a balance between how many ratings we want the user to provide and how much quality we aim to achieve.

In our study, each participant rated one painting from each of the nine categories of the collection we analyzed.
Because the number of categories is small, we could collect one observation from the user for each group of paintings.
However, when the number of categories is too large this approach becomes unfeasible.
To alleviate this, we could explore agglomerative clustering techniques~\cite{rai2010survey}
to select the most interesting groups of paintings to elicit the user's preferences,
based e.g. on dispersion-aware metrics such as cluster intra-variance.

\subsection{Optimizing for real-time performance is important}

We implemented several real-time RecSys engines, where computing performance is critical.
In web applications, it is argued that if users do not receive a response by the system in 1 second,
they will perceive that they do not have control over the system~\cite{Nielsen09}
and quite often they will quit the application if it remains unresponsive~\cite{Hah04}.
To ensure our engines will reply in such a constrained scenario,
we implemented several optimizations, such as using a lightweight version of SBERT
with a small memory footprint instead of the fully-fledged pre-trained model,
and adopting a late fusion technique to merge the contributions of two engines
instead of considering early fusion approaches.

\section{Limitations and future work}

We acknowledge that our crowdsourcing users were not really intrinsically motivated,
or at least not as much as our museum participants, since they had a monetary incentive to take part in the study.
This might have influenced the results, however to mitigate this
we collected a large sample of participants interested in artwork
and considered the user as a random effect in our statistical analysis.

On the other hand, we consider our museum participants intrinsically motivated,
as they were actually visiting a museum, had to scan a QR code with their phones,
and all of them fully completed the study without any monetary compensation.
Also, the correlation coefficient between profile type and recommendation ratings
was higher (and sometimes statistically significant) for museum users.
However, we acknowledge that the sample size is very small to derive general conclusions from that user sample.
The small-scale study, however, agrees with the large-scale crowdsourcing study
in the sense that visual a textual features result in same-quality VA recommendations.
It is therefore advised to consider both approaches when deploying VA RecSys,
as both approaches complement well each other in terms of uncovering different painting semantics.
We believe that, in order to improve the quality of recommendations further,
future work should incorporate more user feedback on artwork, if available
(e.g. in the form of reviews or even the elicited ratings themselves),
as part of our model training pipelines.

As discussed in \autoref{subsec:discuss_one} an interesting takeaway from our study is that the elements of VA recommendation (i.e, key explanatory factors  for semantic relatedness of visual arts) lie not only in visual but also in textual features. We were able to uncover this thanks to our late fusion engines. Particularly the late fusion approach is advantageous as it allows to control the contribution of each fused engine compared to an early fusion approaches in multi-modal feature learning  such as \cite{liu2020dynamic} and a more recent work CLIP \cite{Radford2021} by Open AI. We should note that we used the same backbone architectures as state-of-the-art approaches like CLIP and others [42,58], i.e. Transformers (BERT) for computing text embeddings and ResNet for computing image embeddings. The only difference is that we adopt a late fusion approach since it provides a clear way of understanding the contribution of each modality (image or text) to the generated recommendations. On the contrary, an early fusion approach such as CLIP prevents us from controlling the exact contribution of text and image embeddings because they are entangled, thereby CLIP behaves like a black box model. In our studies, we set exactly 50\% for text and image contribution, respectively.
As a follow-up of this work, we plan to conduct a comparative study of fusion engines (early versus late) on a VA recommendation task.

Finally, we note that our application asked participants to rate one painting
randomly selected from each of the nine categories of our dataset.
This resulted in a 9-dimensional preference elicitation vector with associated weights,
which is perhaps small, considering that previous work asked participants to rate up to 80 paintings~\cite{10.1145/3450613.3456847}.
However, we have not observed substantial overlaps in the rankings produced by each RecSys engine,
which indicates that each participant received truly personalized recommendations.
Further, unlike our experiments, previous work was conducted in a very controlled setting.
In general, preference elicitation is a longstanding challenge in designing real-world RecSys applications.
Ideally, VA RecSys needs to interact with new visitors to gather as much information as possible,
however people are not always willing to provide information or answer lengthy questionnaires~\cite{priyogi2019preference}.
This makes the task of providing personalized VA contents rather challenging.
Hence, instead of relying on explicit user profiling,
future work should investigate efficient strategies to extract maximal information with minimal user engagement.
Nevertheless, we should note that our study reflects a high level of realism, in terms of ecological validity:
anybody can access the application with any device and receive VA recommendations from any of our RecSys engines in real-time.

\section{Conclusion}

Understanding how users’ perceive and interact with highly subjective content such as artwork
is an extremely challenging task due to the complexity of the concepts embedded within artworks
and the emotional and cognitive reflections they may trigger on users.
We have studied the elements of visual art recommendation,
i.e. techniques to uncover latent semantic relationships embedded within paintings,
leveraging textual and visual information, as well as their combination.
To evaluate the performance of each approach, we adopted user-centric evaluation measures.

Our findings open an interesting perspective to understand how users perceive and interact with artwork.
Overall, we can conclude that the semantics of paintings cannot be represented only by visual features nor textual descriptions,
since the emotional and cognitive reflections they may trigger on users are quire diverse and often unpredictable.
Although hybrid approaches of fusing visual and textual features showed clear performance improvements,
more research remains to explore how to improve further the quality of recommendations.

Ultimately, this paper may benefit the HCI community by offering a systematic examination
of how to uncover semantic information from different data sources
in a way that users will perceive as high-quality personalized content.
Our work has potential applications well beyond the scope of this paper,
such as user modeling, intelligent user interfaces, and adaptive user interfaces, among others.
Our dataset, software, and models are publicly available at \url{https://github.com/Bekyilma/VA_RecSys}.

\begin{acks}
This work was supported by the Horizon 2020 FET program of the European Union through the ERA-NET Cofund funding grant CHIST-ERA-20-BCI-001
and the European Innovation Council Pathfinder program (SYMBIOTIK project, grant 101071147).
\end{acks}

\balance
\bibliographystyle{ACM-Reference-Format}
\bibliography{main}

\end{document}